\documentclass[english,british,nofootinbib]{revtex4-2}
\PassOptionsToPackage{obeyFinal}{todonotes}
\usepackage[T1]{fontenc}
\usepackage[latin9, utf8]{inputenc}
\usepackage{color}
\usepackage{todonotes}
\usepackage{amsmath}
\usepackage{amssymb}
\usepackage{graphicx}
\usepackage{esint}



\usepackage{babel}
\begin{document}
\selectlanguage{english}%
\global\long\def\reh{\mathrm{reh}}%
\foreignlanguage{british}{}
\global\long\def\mpl{m_{\mathrm{Pl}}}%
\foreignlanguage{british}{ }
\global\long\def\c{\mathrm{SBP}}%
\foreignlanguage{british}{ }
\global\long\def\i{\mathrm{i}}%

\title{Primordial Black Holes as Dark Matter and the Tachyonic Trap During Inflation}
\author{Yuma S. Furuta}
\affiliation{School of High Energy Accelerator Science, Graduate University for Advanced Studies (SOKENDAI), 1-1 Oho, Tsukuba, Ibaraki 305-0801, Japan}
\affiliation{Theory Center, IPNS, High Energy Accelerator Research Organization (KEK), 1-1 Oho, Tsukuba, Ibaraki 305-0801, Japan}
\author{Mindaugas Kar\v{c}iauskas}
\affiliation{Center for Physical Sciences and Technology (FTMC), Saul\.{e}tekio av. 3, 10257 Vilnius, Lithuania}
\author{Kazunori Kohri}
\affiliation{Division of Science, National Astronomical Observatory of Japan, 2-21-1 Osawa, Mitaka, Tokyo 181-8588, Japan}
\affiliation{School of Physical Sciences, Graduate University for Advanced Studies (SOKENDAI), 2-21-1 Osawa, Mitaka, Tokyo 181-8588, Japan}
\affiliation{Theory Center, IPNS, High Energy Accelerator Research Organization (KEK), 1-1 Oho, Tsukuba, Ibaraki 305-0801, Japan}
\affiliation{Kavli IPMU (WPI), UTIAS, The University of Tokyo, Kashiwa, Chiba 277-8583, Japan}
\author{Alejandro Sáez}
\address{Instituto de Física Corpuscular (IFIC), CSIC-Universitat de València, 46071, Valencia, Spain}

\begin{abstract}
    We show that resonant processes during multi-field inflation can generate 
    a large curvature perturbation on small scales. 
    This perturbation naturally leads to the formation of primordial black holes 
    that may constitute dark matter, as well as to
    the production of stochastic induced gravitational waves in the deci-Hz band.
    Such waves are within reach of future space-based interferometers such
    as LISA, DECIGO and BBO. In addition, primordial black hole binaries formed 
    at late times produce merger gravitational waves that can be probed by the 
    resonant cavity experiments in addition to DECIGO and BBO.
\end{abstract}
\maketitle

\section{Introduction}\label{sec:Intro}

In several multi-field inflationary scenarios, the Universe undergoes
multiple-field transitions during inflation, analogous to phase
transitions. Identifying the role of these transitions through
observations could provide decisive information about the ultimate
theory to describe the early Universe.

In some of these models, interactions between fields can cause resonant field 
excitations during inflation. 
A classical example is the so called ``Trapped Inflation'' \cite{Kofman:2004yc,Green:2009ds}.
But resonances during inflation and their impact on the primordial perturbation 
had been studied even earlier, for example, in Ref.
~\cite{Chung:1999ve,Elgaroy:2003hp,Romano:2008rr} (for some later works see 
Ref.~\cite{Barnaby:2009mc,Barnaby:2010sq,Pearce:2017bdc,Cai:2021yvq}). 
Many of the works are based on the mechanism suggested in 
Ref.~\cite{Kofman:2004yc}, where moduli fields of string theory are stabilised by the 
backreaction of resonantly produced particles. 
As moduli field(s) pass through or close to (in the case of multi-field inflation) 
a critical point in field space, some real scalar field $\chi$ is rendered massless. 
Thus, this point was naturally denoted by the name ``Enhanced Symmetry Point'' (ESP) in Ref.~\cite{Kofman:2004yc}.
Around ESP the effective mass of $\chi$ is changing non-adiabatically,
which induces resonant excitations.
The excitations backreact onto the motion of the inflaton, modifying its 
dynamics.

In this work we consider a somewhat modified scenario. 
Instead of the field $\chi$ becoming massless at the critical point, we allow 
it to become tachyonic, i.e. it's mass squared to become negative. Hence, for
some parameter values the resonance resembles the one studied in Ref.~\cite{Dufaux:2006ee},
which they called ``Tachyonic Resonance''.
To emphasise this difference, we name the critical point as the ``Symmetry 
Breaking Point'' (SBP). 
The idea for such a scenario is inspired by the tachyonic trap 
mechanism, employed to provide an alternative method of reheating for
non-oscillatory potentials and to prevent the scalar field of the quintessential inflation
scenario from reaching superplanckian values \cite{Dimopoulos:2019ogl,Karciauskas:2021fdu}. 

The current scenario is explored in the context of the supersymmetry-inspired 
multi-field running-mass model~\cite{Leach:2000ea}. 
Augmenting the running-mass-inflation (RMI) with the tachyonic trap mechanism 
provides a concrete scenario by which inflation can end in a form similar
to hybrid inflation \cite{Linde:1993cn,Copeland:1994vg}. 
The current treatment, however, differs from the usual considerations of hybrid
inflation. 
First, we take the waterfall phase, i.e. the evolution in the $\chi$ direction
in field space, to last more than 10 e-folds, which requires a very flat potential.
In such a setup, the tachyonic trap becomes essential to redirect the field 
evolution from the RMI direction into the waterfall one, at least for parameter
ranges considered in this work.

The tachyonic resonance at SBP has another important effect: it generates a
sharp peak in the spectrum of the primordial perturbation ${\cal P}_{\zeta}$
at small scales.
The amplitude of that peak can be enhanced by several orders of magnitude 
relative to the slow-roll result. 
Combined with the large spectral running of RMI, the spectrum can reach as high 
values as ${\cal P}_{\zeta} \sim 10^{-1.5}$~\cite{Leach:2000ea,Kohri:2007qn,Alabidi:2009bk,Alabidi:2012ex,Alabidi:2013lya,Inomata:2017okj,Inomata:2016rbd,Kohri:2018qtx}.

Later, upon horizon reentry in the early Universe, the large curvature
perturbation can trigger gravitational collapse, leading to the
formation of the primordial black holes (PBHs). 
As argued in Ref.~\cite{Carr:2020gox} (for reviews, see also
Refs.~\cite{Carr:2020xqk,Green:2020jor,Escriva:2022duf}), PBHs with
masses in the range $10^{17}$~g~\cite{Carr:2009jm} --
$10^{23}$~g~\cite{Niikura:2017zjd,Smyth:2019whb} can serve as dark
matter candidates. This scenario can be constrained or confirmed by
future gamma-ray observations and related astrophysical
probes~\cite{Carr:2020gox}.

Second, the same perturbations generate stochastic induced
gravitational waves (SIGWs) through nonlinear second-order
effects~\cite{Mollerach:2003nq,Ananda:2006af,Baumann:2007zm,Saito:2008jc,Bugaev:2009zh,Assadullahi:2009nf,Espinosa:2018eve,Kohri:2018awv,Cai:2019cdl}. These
SIGWs, peaking in the deci-Hz band, provide promising targets for
future gravitational-wave detectors such as LISA, DECIGO, and BBO. In
the current analysis, we incorporate recent refinements accounting for
the dissipation of small-scale fluctuations~\cite{Domenech:2025bvr},
yielding state-of-the-art predictions for the SIGW spectrum.

Third, PBHs formed in the early Universe can assemble into binaries
through many-body gravitational
interactions~\cite{Sasaki:2018dmp,Wang:2019kaf,Kohri:2024qpd}. The
mergers of these binaries also generate a stochastic background of
gravitational waves. Remarkably, such merger signals may be detectable
not only by DECIGO and BBO but also through resonant cavity
experiments exploiting the inverse Gertsenshtein
effect~\cite{Berlin:2021txa, Herman:2022fau}, originally proposed in
the context of axion searches.

The remainder of this paper is organized as follows. In Sec. II, we
outline the general framework of trapped inflation. Sec. III applies
it to the running-mass model in supersymmetry. Sec. IV describes the
trapping mechanism in details, and Sec. V quantifies the trapping
duration during the second stage of inflation. Sec. VI reviews the
basics of primordial curvature perturbations, and Sec. VII derives the
PBH mass function. Sec. VIII presents the calculation of induced
gravitational waves, while Sec. IX addresses gravitational waves from
PBH binary mergers. Sec. X summarizes our conclusions. Technical
details are collected in three appendices: in Appendix A, Primordial
perturbations in the flat gauge, in Appendix B, Stochastic induced GWs, and, in Appendix C, Binary PBH merger GWs. 
We use the units where $c=\hbar=1$, $\mpl=\left(8\pi G\right)^{-1/2}$
and $G$ is the Newton's gravitational constant.

\section{The Model}\label{sec:model}
\begin{figure}
\begin{centering}
\includegraphics[scale=0.7]{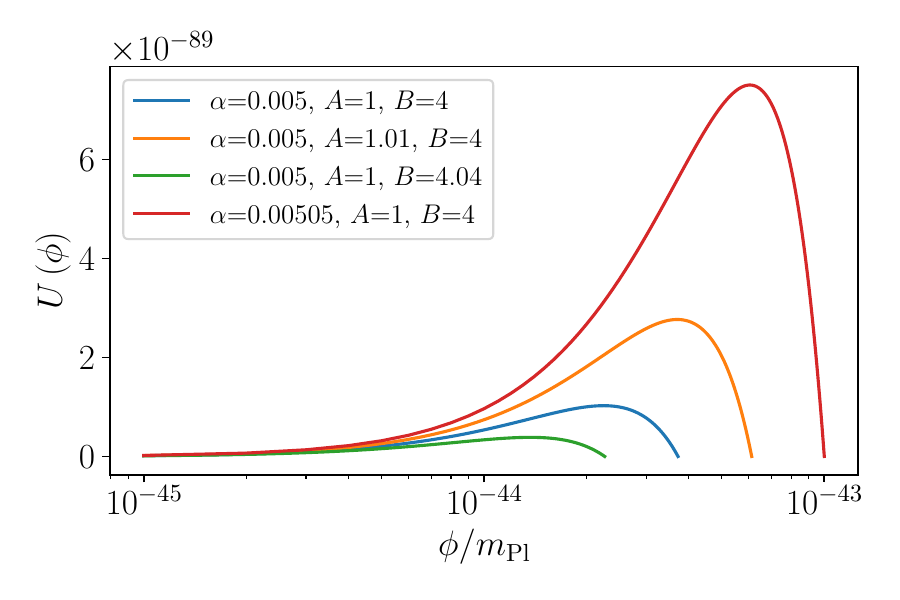}
\par\end{centering}
\caption{\label{fig:U}Rescaled RMI potential $U\left(\phi\right)$ in
Eq.~(\ref{Udef}) for several $\alpha$, $A$ and $B$ values.}

\end{figure}

We study a multi-field model of inflation. The basic setup is
reminiscent of the hybrid inflation scenario \citep{Linde:1993cn,Copeland:1994vg}
with some essential modifications. First, we consider the potential in
the direction of the inflaton $\phi$ to be completely flat at
tree level. The slope is generated by radiative corrections. Such models are 
known by the name "Running-Mass-Inflation"
(RMI) \citep{Stewart:1996ey,Stewart:1997wg,Covi:1998jp}. At one loop
level, the shape of the potential is given by
\begin{eqnarray}
V\left(\phi\right) & = & V_{\mathrm{c}}\left[1+U\left(\phi\right)\right]\label{rmiV}
\end{eqnarray}
where $V_{\mathrm{c}}$ is constant and $U\left(\phi\right)$ is defined
by
\begin{eqnarray}
U\left(\phi\right) & \equiv & -\frac{1}{2}\frac{\phi^{2}}{\mpl^{2}}\left(B-\frac{A}{\left(1+\alpha\ln\frac{\phi}{\mpl}\right)^{2}}\right)\label{Udef}
\end{eqnarray}
The shape of $U\left(\phi\right)$ for several values of $\alpha$,
$A$ and $B$ parameters is shown in Fig.~\ref{fig:U}. 

In analogy to hybrid inflation scenario we add an additional field
$\chi$. The potential in the $\chi$ direction is of the hilltop
type \citep{Lyth:2009zz}. The simplest form of such a potential can
be written as 
\begin{eqnarray}
V\left(\chi\right) & = & -\frac{1}{2}m^{2}\chi^{2}+\lambda\chi^{4}\,.\label{inf2V}
\end{eqnarray}

The crucial piece for this scenario is the interaction part of the
Lagrangian, which we take to be \citep{Dimopoulos:2019ogl,Karciauskas:2021fdu}
\begin{eqnarray}
V_{\mathrm{int}}\left(\phi,\chi\right) & = & \frac{1}{2}g^{2}\chi^{2}\left(\phi-\phi_{\c}\right)^{2}\,.\label{tachtrap}
\end{eqnarray}
$\phi_{\c}$ denotes the critical value at which $V_{\mathrm{int}}$
vanishes. We choose to call this point a "Symmetry Breaking Point" and denote 
it by $\phi_{\c}$ in order to make the relation to the mechanism discussed in 
Ref.~ \cite{Karciauskas:2021fdu} more suggestive, but also to emphasise that our
scenario somewhat deviates from the standard hybrid inflation scenario.

The main role of the $\chi$ field in hybrid inflation scenarios is
to terminate inflation. It is called the waterfall field. In our scenario
the main function of $\chi$ is to trap $\phi$ at the $\phi_{\c}$
value. This is achieved by the backreaction of resonantly produced $\chi$ 
particles \cite{Kofman:2004yc,Karciauskas:2021fdu}, hence we also sometimes call $\chi$ as the ``trapping field''. 
Another difference, as compared to the traditional hybrid scenario, is that
the ``waterfall'' phase in this scenario lasts many e-folds. This is needed to 
extend inflation sufficiently long after $\phi=\phi_{\c}$ is reached, so that the
horizon and flatness problems of Hot Big Bang (HBB) are solved. Finally, as can
be witnessed from Eq.~\eqref{tachtrap}, the $\phi$--$\chi$ interaction includes
a trilinear term. Trilinear interactions can be found in the $A$ term of SUGRA models
\cite{Kachru:2003sx}, but it can also be generated by fermion condensation \cite{Iso:2014gka}.

Adding all these components together, the full Lagrangian of the model can
be written as
\begin{eqnarray}
\mathcal{L} & = & -\frac{1}{2}\left(\partial_{\mu}\phi\right)^{2}-V\left(\phi\right)-\frac{1}{2}\left(\partial_{\mu}\chi\right)^{2}-V\left(\chi\right)-V_{\mathrm{int}}\left(\phi,\chi\right)\label{L}
\end{eqnarray}

The dynamics evolves over three stages. Initially the inflaton $\phi$
is displaced far away from the critical value, $\phi\gg\phi_{\c}$.
This makes the trapping field $\chi$ very heavy and anchored at the
origin. During the first phase, while $\phi$ rolls down towards the
origin, the dynamics can be well approximated by slow-roll. Once
$\phi$ approaches $\phi_{\c}$ the second phase starts. The motion
of $\phi$ induces a non-adiabatic change in the effective mass of
the trapping field via the interaction term in Eq.~(\ref{tachtrap}).
This results in resonant excitations of $\chi$, which backreact onto
the motion of $\phi$ and anchors it at $\phi_{\c}$. In the final
phase, the $\chi$ field rolls down the potential in Eq.~(\ref{inf2V}),
which is chosen to be sufficiently flat, so that inflation lasts for
an additional $\sim10$ e-folds in this phase. Bellow we discuss these
phases in more detail.

\section{Running-Mass-Inflation}\label{sec:RMI}
Let $\phi_{*}$ be the inflaton field value when the pivot scale exits
the horizon during inflation. CMB observations allow us to constrain
the primordial spectrum roughly 10 e-folds around this value. The
first task is to find regions in $\left(\alpha,A,B\right)$ parameter space
where the model generates the primordial perturbation
that is consistent with CMB observations. 

Generically we take $\phi_{\c}\ll\phi_{*}$, which, according to Eq.~(\ref{tachtrap}),
makes the $\chi$ field heavy and anchored at the origin, leading
to an effectively single field inflation, at least within the 10 e-folds
mentioned above. Another important consequence of $\chi$ being heavy
is that the isocurvature perturbation is suppressed at CMB scales, which makes 
the observational bounds on this parameter \citep{Planck:2018jri} easily satisfied.

\begin{figure}
\begin{centering}
\includegraphics[scale=0.7]{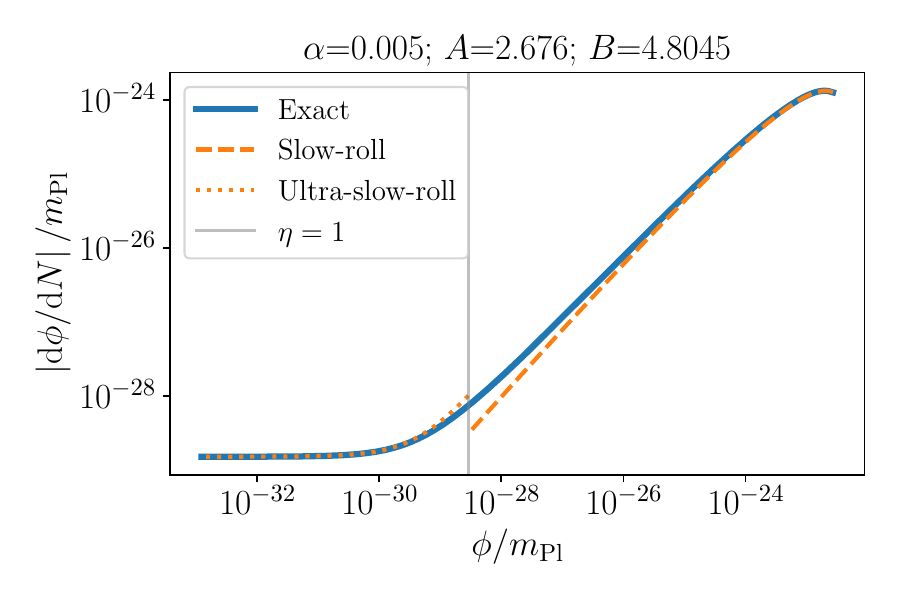}
\par\end{centering}
\caption{\label{fig:phase}The numerical solution of homogeneous RMI equations (``Exact'').
At the pivot scale, on the RHS of the plot, inflation is well approximated
by slow-roll (cf. Eq.~(\ref{sr})), but eventually it enters the ultra-slow-roll regime (cf. Eq.~(\ref{usr})). Conventionally
this transition is taken at $\eta=1$ (vertical gray line).
}

\end{figure}

One of the features of the RMI potential in Eq.~(\ref{rmiV}) is
that it becomes ever flatter as $\phi$ field approaches the origin.
This makes the inflaton dynamics eventually dominated by the kinetic
energy rather than by the slope of the potential. In other words, inflation
enters the ultra-slow-roll regime \citep{Kinney:1997ne,Inoue:2001zt,Kinney:2005vj,Martin:2012pe}
if not terminated earlier. To investigate this issue we solved the homogeneous
equations of motion numerically. One such solution is shown in Fig.~\ref{fig:phase}.

As can be seen from Fig.~\ref{fig:phase}, at larger inflaton values slow-roll
provides a good description of the dynamics. During this period the
inflaton equation of motion can be approximated by
\begin{eqnarray}
\dot{\phi} & \simeq & -\frac{V_{,\phi}}{3H}\,,\label{sr}
\end{eqnarray}
where $V$ is given in Eq.~(\ref{rmiV}) and the index denotes the
derivative with respect to the field $\phi$. In this approximation
the Hubble parameter is dominated by the potential energy
\begin{eqnarray}
    3\mpl^{2}H^{2} & \simeq & V\left(\phi\right)\simeq V_{\mathrm{c}}\,,\label{H-apx}
\end{eqnarray}
where $V_\mathrm{c}$ is defined in Eq.~\eqref{rmiV}. 
This slow-roll approximated solution is denoted by the dashed curve
in Fig.~\ref{fig:phase}.

As $\phi$ decreases, eventually inflation enters the ultra-slow-roll
regime. In this regime the slope of the potential can be neglected,
and we obtain an approximate equation of motion of the form
\begin{eqnarray}
\ddot{\phi} & \simeq & -3H\dot{\phi}\,,
\end{eqnarray}
where $H\simeq\mathrm{const}.$ and its value can be computed using
the same approximate expression in Eq.~(\ref{H-apx}). It is easy
to show that the approximate solution of the above equation is
\begin{eqnarray}
\dot{\phi} & \simeq & \dot{\phi}_{0}-3H\left(\phi-\phi_{0}\right)\,.\label{usr}
\end{eqnarray}
The ultra-slow-roll approximated solution is represented by the dotted curve in Fig.~\ref{fig:phase}. 

The approximate location in the potential, where slow-roll gives way
to ultra-slow-roll, is conventionally taken to be $\eta=1$, where
$\eta$ is the second slow-roll parameter defined bellow in Eq.~(\ref{eta-def}).

To choose viable models, which do not contradict CMB constraints,
we calculate the properties of the scalar perturbation spectrum and
the amplitude of the tensor mode. Because slow-roll approximates the
inflaton dynamics sufficiently well when CMB scales exit the horizon,
we use the well known relations between the slow-roll parameters and
the shape of the primordial spectrum. These parameters are defined
in terms of the potential $V\left(\phi\right)$ and its derivatives
as
\begin{eqnarray}
\epsilon & = & \frac{\mpl^{2}}{2}\left(\frac{U_{,\phi}}{U+1}\right)^{2}\,,\label{epsilon-def}\\
\eta & = & \mpl^{2}\frac{U_{,\phi\phi}}{U+1}\,,\label{eta-def}\\
\xi^{2} & = & \mpl^{4}\frac{U_{,\phi}U_{,\phi\phi\phi}}{\left(U+1\right)^{2}}\,,\\
\omega^{3} & = & \mpl^{6}\frac{U_{,\phi}^{2}U_{,\phi\phi\phi\phi}}{\left(U+1\right)^{3}}\,,
\end{eqnarray}
where $U\left(\phi\right)$ is given in Eq.~(\ref{Udef}) and the
indices denote derivatives with respect to the inflaton $\phi$. The
spectral properties of the primordial scalar perturbation are related
to the above parameters by the following expressions (see e.g. Ref.~\citep{Lyth:2009zz}
or \citep{Planck:2013jfk})
\begin{eqnarray}
n_{\mathrm{s}}\equiv\frac{\mathrm{d}\ln A_{\mathrm{s}}}{\mathrm{d}\ln k} & = & 1-6\epsilon+2\eta\,,\\
\alpha_{\mathrm{s}}\equiv\frac{\mathrm{d}\ln n_{\mathrm{s}}}{\mathrm{d}\ln k} & = & 16\epsilon\eta-24\epsilon^{2}-2\xi^{2}\,,\\
\beta_{\mathrm{s}}\equiv\frac{\mathrm{d}^{2}\ln n_{\mathrm{s}}}{\mathrm{d}\ln k^{2}} & = & -192\epsilon^{3}+192\epsilon^{2}\eta-32\epsilon\eta^{2}-24\epsilon\xi^{2}+2\eta\xi^{2}+2\omega^{3}\,,
\end{eqnarray}
where $n_{\mathrm{s}}$, $\alpha_{\mathrm{s}}$ and $\beta_{\mathrm{s}}$ are the 
scalar spectral index, its running and the running-of-the-running respectively. 
All of these quantities are to be computed when the pivot scale exits the horizon. 
Analogously, tensor-to-scalar ratio can also be related to the slow-roll
parameter $\epsilon$ by
\begin{eqnarray}
r & = & 16\epsilon\,.
\end{eqnarray}

\begin{figure}
\begin{centering}
\includegraphics[scale=0.5]{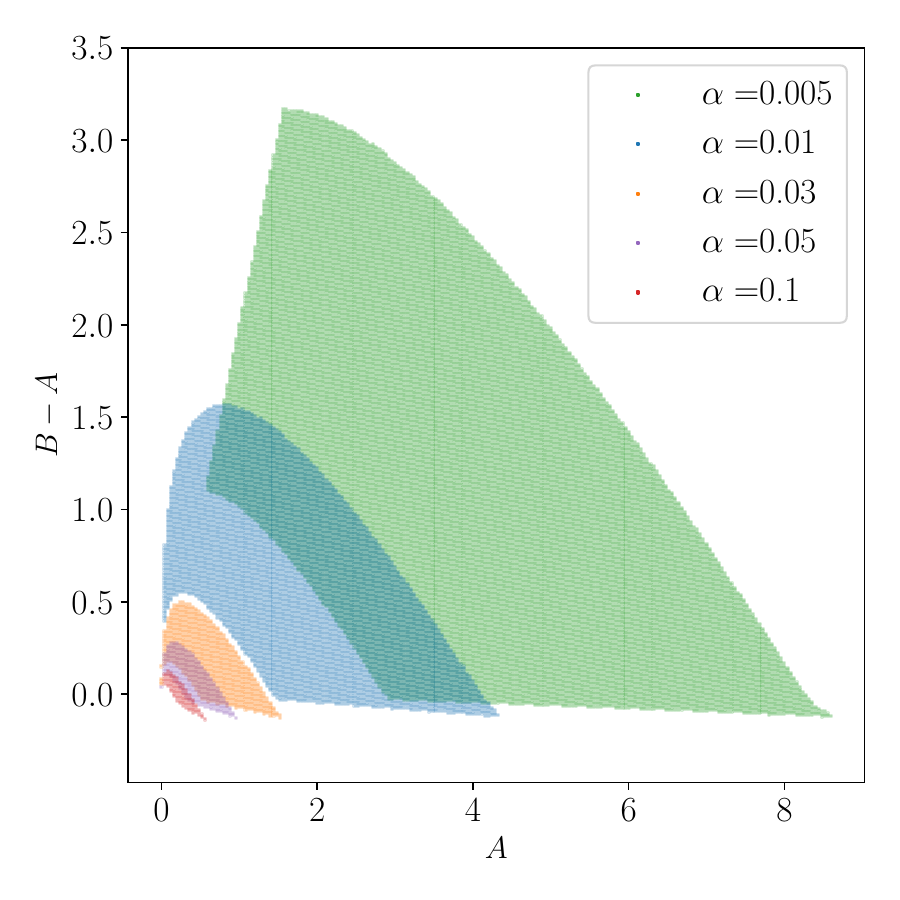}
\par\end{centering}
\caption{\label{fig:inf-constr}
    Parameter regions for several different $\alpha$
    values where the running mass inflation models are compatible with
    the CMB constraints on the primordial spectrum at the pivot scale
    $k_*=0.05\,\mathrm{Mpc}^{-1}$ (Eqs.~(\ref{ns-const})--(\ref{r-const})). 
     We also impose two other
    conditions: $\phi_{*}<\protect\mpl$ when the pivot scale exits the horizon
    and that the energy scale of inflation is larger than the Big Bang 
    Nucleosynthesis (BBN) scale.
}
\end{figure}

To find models that are compatible with observations we scan over
the parameters $\alpha$, $A$ and $B$ and look for regions of $\phi_{*}$
values that result in the primordial spectrum with values in the range
\begin{eqnarray}
n_{s} & = & 0.9743\pm0.0034\,,\label{ns-const}\\
\alpha_{s} & = & 0.0062\pm0.0052\,,\label{as-const}\\
\beta_{s} & = & 0.010\pm0.013\,.\label{bs-const}
\end{eqnarray}
The errorbars correspond to $1\sigma$ constraints for $n_s$ \citep{ACT:2025fju},
$\alpha_s$~\citep{ACT:2025tim} and $\beta_s$~\citep{Planck:2018jri}.
To constrain the tensor-to-scalar ratio we adopt the
upper bound in Ref.~\citep{BICEP:2021xfz}
\begin{eqnarray}
r & < & 0.036\,.\label{r-const}
\end{eqnarray}
The final PBH abundance is not very sensitive to the precise values of these 
parameters. They only determine which model in the $\left(\alpha,A,B\right)$
plane will be used to represent the inflaton direction.
The result of the scan over this parameter space is shown in 
Fig.~\ref{fig:inf-constr} for several values of $\alpha$.

Usually all the constraints in Eqs.~(\ref{ns-const})--(\ref{r-const})
can be satisfied only for a small range of $\phi$ values, if at all.
For such models we choose $\phi_{*}$ to correspond to the value that
is closest to the central value of the constraints. Once $\phi_{*}$
is fixed, we can compute the energy scale of inflation $V_{\mathrm{c}}$
from the amplitude of the scalar spectral index $A_\mathrm{s}$, given by
\begin{eqnarray}
A_{\mathrm{s}} & = & \frac{V}{24\mpl^{2}\epsilon}\,.
\end{eqnarray}
The value of $A_{\mathrm{s}}$ is fixed by the Planck normalisation,
$A_{\mathrm{s}}=3.044$ ~\citep{Planck:2018jri}.

Initially $\phi$ follows the slow-roll equation of motion in Eq.~\eqref{sr}.
As the potential flattens out the dynamics becomes well approximated
by ultra-slow-roll in Eq.~(\ref{usr}). The latter equation is solved
by 
\begin{eqnarray}
\dot{\phi} & \propto & a^{-3}\,.
\end{eqnarray}

One of the consequences of the flattening of the potential is the
rapid increase in the amplitude of the curvature perturbation $\mathcal{P}_{\zeta}$.
On superhorizon scales it can be written as
\begin{eqnarray}
\mathcal{P}_{\zeta}\left(k\right) & = & \left(\frac{H}{\dot{\phi}}\right)^{2}\left|\delta\phi_{k}\right|^{2}\,,\label{Pz}
\end{eqnarray}
where $\delta\phi_{k}\left(t\right)$ is the Fourier mode of the field
perturbation $\delta\phi\left(x,t\right)$. We can see that as $\dot{\phi}$
decreases, $\mathcal{P}_{\zeta}$ rapidly grows. We must make sure
that the first phase of inflation is terminated before $\mathcal{P}_{\zeta}$ reaches
the value of 1. Otherwise perturbations become non-linear, which is
in conflict with observations \citep{Bringmann:2011ut,Inomata:2018epa}.

\section{The Trapping Phase}\label{sec:trap}
To the best of our knowledge there is no detailed discussion in the
literature of a mechanism to end the RMI phase and provide the remaining
e-folds of inflation. Usually it is implicitly assumed that hybrid
inflation or some related mechanism terminates the RMI stage before $\mathcal{P}_{\zeta}\left(k\right)$
becomes too large. Unfortunately, a simplistic implementation of hybrid
inflation is difficult to realise. After observable scales -- where 
$\mathcal{P}_{\zeta}\left(k_*\right)$ is fixed to satisfy CMB bounds -- exit the horizon,
the maximum value of $\mathcal{P}_{\zeta}\left(k\right)$ must be reached in about 30-35 e-folds.
Only then the masses of PBHs created by a large curvature perturbation are such that 
they can explain the observed DM abundance (see Fig.~\ref{fig:fpbh}). 
But 30-35 fall short from the expected 50-60 e-folds of inflation, which are 
required to solve the flatness and horizon problems of HBB (see section \ref{duration} for a more detailed discussion).
Therefore, to solve these problems inflation must last for an additional 
$\sim10$ e-folds in the waterfall phase.
A long waterfall phase can be achieved if the potential in this direction is flat enough. 
But we found that for a too flat potential the $\phi$ field just 
zips through the critical (Symmetry Breaking) point without destabilising the 
waterfall field $\chi$. 

\begin{figure}
\begin{centering}
\includegraphics[scale=0.4]{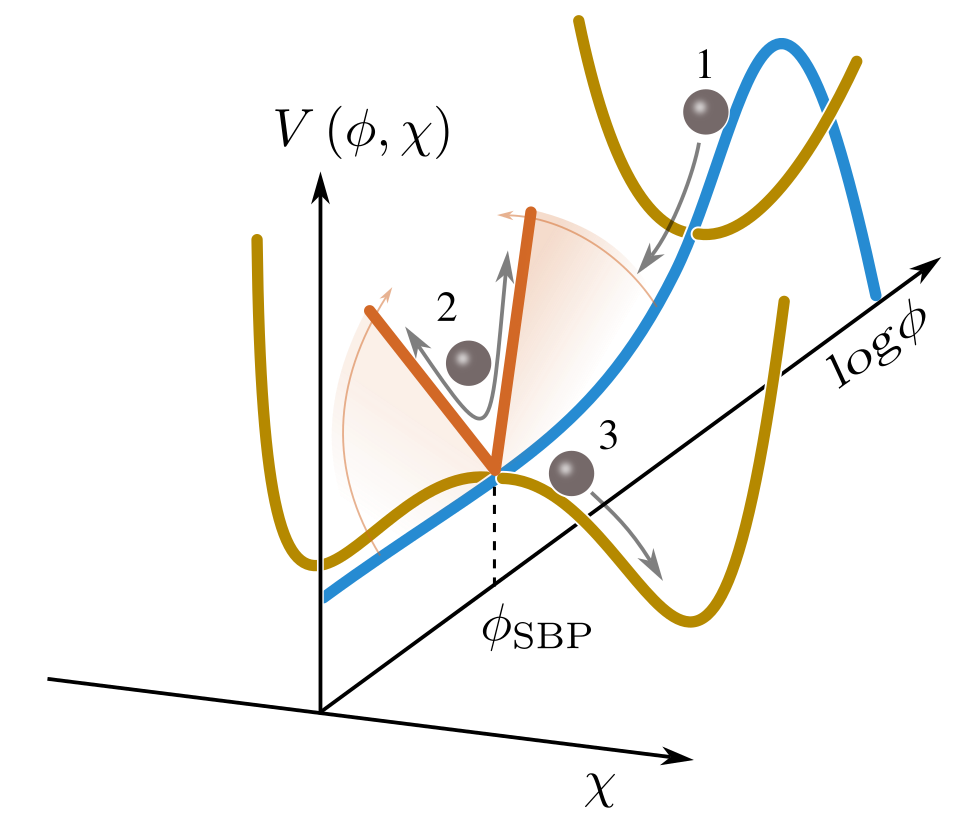}
\par\end{centering}
\caption{\label{fig:scenario}
    Schematic depiction of our scenario. At stage
    1 the field rolls down along the RMI direction (blue curve). At tree
    level, the linear potential (red curve) does not exist. Once the field
    reaches SBP (stage 2) it resonantly excites the $\chi$ field. Excitations
    backreact onto the motion of the field, which can be effectively described
    by a steepening linear potential. At this stage the trapped $\phi$ field
    oscillates around $\phi_{\protect\c}$ with an exponentially decreasing
    amplitude. Eventually the amplitude becomes too small to excite the
    $\chi$ field. At that point the evolution enters the 3rd stage, in
    which the field rolls down solely in the direction of the sufficiently flat
    waterfall potential (brown curve) with $\phi$ remaining being fixed at $\phi_\c$. 
    During stage 2 the metric perturbation is also resonantly amplified for 
    scales which exit the horizon at that time.}
\end{figure}

Fortunately, as we show, this problem can be circumvented by another effect.
For some parameter values the passage of $\phi$ through the Symmetry Breaking Point $\phi_{\c}$
induces resonant excitations of the $\chi$ field. This field backreacts
onto the motion of $\phi$ by making it effectively heavy and stopping it from rolling
down the potential. This is the basic scenario of the tachyonic trap
mechanism discussed in Refs.~\citep{Dimopoulos:2019ogl,Karciauskas:2021fdu}.
In this work we make use of the tachyonic trap mechanism to terminate
RMI at the value of $\mathcal{P}_{\zeta}$ that results in the production
of PBHs with masses that can explain Dark Matter \citep{Carr:2020gox}.
A schematic depiction of our scenario is provided in Fig.~\ref{fig:scenario}.

\subsection{The Tachyonic Trap\label{sec:tachtrap}}
In Refs.~\citep{Dimopoulos:2019ogl,Karciauskas:2021fdu}, where the tachyonic
trapping mechanism is analysed, the metric perturbations are ignored. 
These simplifications can no longer be employed for the current model, where 
such perturbations play the central role. 
Nevertheless, before discussing the model in full detail, including the metric 
perturbation, bellow we summarize the basic ideas behind the tachyonic trap mechanism.

For the most part of the RMI stage of inflation the trapping field
$\chi$ is very heavy. This is the case if the value of the coupling
constant $g$ in Eq.~\eqref{tachtrap} is not too small, so that
$g^{2}\left(\phi-\phi_{\c}\right)^{2}\gg H^{2}$, where $H$ is the
Hubble parameter during inflation. This makes the homogeneous component
of the $\chi$ field anchored at the origin. The perturbations $\delta\chi$
of the $\chi$ field obey the following equation of motion: 
\begin{eqnarray}
\delta\ddot{\chi}_{k}+3H\delta\dot{\chi}_{k}+\left(\omega_{k}^{2}+12\lambda\left\langle \chi^{2}\right\rangle \right)\delta\chi_{k} & \simeq & 0\,,\label{dchi-eom-nog}
\end{eqnarray}
where
\begin{eqnarray}
\omega_{k}^{2} & \equiv & \frac{k^{2}}{a^{2}}-m^{2}+g^{2}\left(\phi-\phi_{\c}\right)^{2}
\end{eqnarray}
We can rewrite the above equation in the canonical form by defining
\begin{eqnarray}
X_{k} & \equiv & a\chi_{k}
\end{eqnarray}
and using the conformal time $\mathrm{d}\tau\equiv\mathrm{d}t/a$.
This gives
\begin{eqnarray}
X_{k}''+\left(W_{k}^{2}+12\lambda a^{2}\left\langle \chi^{2}\right\rangle \right)X_{k} & = & 0\,,\label{X-eom}
\end{eqnarray}
where primes denote derivatives with respect to $\tau$ and
\begin{eqnarray}
W_{k}^{2} & \equiv & a^{2}\omega_{k}^{2}-\frac{a''}{a}\,.
\end{eqnarray}

Initially, as field $\phi$ is far away from SBP, the effective mass of the trapping field satisfies
\begin{equation}
m_{\mathrm{eff}}^{2}\equiv g^{2}\left(\phi-\phi_{\c}\right)^{2}-m^{2}\simeq g^{2}\left(\phi-\phi_{\c}\right)^{2}\gg12\lambda\left\langle \chi^{2}\right\rangle
\end{equation}
and Eq.~\eqref{X-eom} reduces to the equation of a harmonic oscillator with an adiabatically changing mass. 
We can thus impose the adiabatic vacuum initial conditions, which, at the 
lowest order, are given by
\begin{eqnarray}
X_{k,\mathrm{vac}}\left(\eta\right) & = & \frac{1}{\sqrt{2W_{k}}}\mathrm{e}^{-\mathrm{i}\int W_{k}\mathrm{d}\eta}\,.
\end{eqnarray}
Up to the same order, the occupation number can be computed using
the expression 
\begin{eqnarray}
n_{k} & = & \frac{W_{k}}{2}\left[\frac{\left|X_{k}'\right|^{2}}{W_{k}^{2}}+\left|X_{k}\right|^{2}\right]-\frac{1}{2}\,.
\end{eqnarray}

The $12\lambda a^{2}\left\langle \chi^{2}\right\rangle $
term consists of the Hartree approximation to account for self-interactions,
where the expectation value $\left\langle \chi^{2}\right\rangle $
can be computed using the equation
\begin{eqnarray}
    \left\langle \chi^{2}\right\rangle =a^{-2}\left\langle X^{2}\right\rangle  & = & \frac{a^{-2}}{2\pi^{2}}\int \mathrm{d}k\, k^{2}\left[\left|X_{k}\right|^{2}-\frac{1}{2\left|W_{k}\right|}\right]\,.\label{chi2}
\end{eqnarray}

As the $\phi$ field moves towards the origin and comes close to $\phi_{\c}$
the effective mass squared $m_{\mathrm{eff}}^{2}$ vanishes and then
becomes negative. Moreover, within some interval of $\phi$ values
the change of $m_{\mathrm{eff}}^{2}$ is rendered to be non-adiabatic~\cite{Dimopoulos:2019ogl,Karciauskas:2021fdu}.
This causes two effects. First, the non-adiabaticity of $m_{\mathrm{eff}}^{2}$
results in the resonant excitations of the $\chi$ field, as described
in Ref.~\citep{Kofman:1997yn}. Second, as $m_{\mathrm{eff}}^{2}$
becomes negative, it can lead to an additional amplification of $\chi$
field perturbations via the process known as the tachyonic resonance
~\citep{Felder:2001kt,Dufaux:2006ee}. 

Which of the two effects dominates, depends on model parameters \citep{Karciauskas:2021fdu}.
But in both cases, due to $\phi$--$\chi$ interactions, the exponential
growth of $\left\langle \chi^{2}\right\rangle $ backreacts onto the
motion of the $\phi$ field by creating an effective contribution
to its mass term. Indeed, from Eq.~(\ref{tachtrap}) we find the
effective equation of motion of the homogeneous component of the $\phi$
field to be
\begin{eqnarray}
\ddot{\phi}+3H\dot{\phi}+V\left(\phi,\chi\right)_{,\phi} & = & 0\,,
\end{eqnarray}
where 
\begin{eqnarray}
V\left(\phi,\chi\right)_{,\phi} & \simeq & V_{\mathrm{c}}U_{,\phi}+g^{2}\left\langle \chi^{2}\right\rangle \left(\phi-\phi_{\c}\right)
\end{eqnarray}
and $V_{\mathrm{c}}$ and $U$ are defined in Eqs.~(\ref{rmiV})
and (\ref{Udef}) respectively. Once the second term in the above
expression becomes large enough, $g^{2}\left\langle \chi^{2}\right\rangle >H^{2}$,
the field becomes heavy and stops running towards the origin, but
rather oscillates around $\phi_{\c}$ with a decaying amplitude.

This process is somewhat similar to the one described in Ref.~\citep{Kofman:2004yc},
where the resonant excitations of the $\chi$ field traps the $\phi$ field at SBP. 
In contrast to that work, we take $\chi$ to be tachyonic. 
This way the trapping of $\phi$ at
$\phi_{\c}$ initiates the symmetry breaking phase by releasing the
$\chi$ field from the origin and allowing it to roll towards the
vacuum value. This gives the name for the subscript $\phi_{\c}$,
as in ``Symmetry Breaking Point'' and the name ``tachyonic trap''
for the mechanism \citep{Karciauskas:2021fdu}.

There is another crucial difference in the current model as compared 
to both Ref.~\citep{Kofman:2004yc} and \citep{Karciauskas:2021fdu}.
The resonance and the trapping in the latter references are assumed to
happen in a non-accelerating spacetime. In the current application,
we make use of the tachyonic trap during inflation. The idea of resonant
field excitations during inflation is not new. We can find such discussions
in, for example, Refs.~\citep{Chung:1999ve,Battefeld:2011yj,Pearce:2017bdc,Naik:2022mxn}
and many others. But here again, in contrast to those works, we consider
$\chi$ to be tachyonic. This allows for the $\chi$ field to play
the role of the waterfall field à la hybrid inflation.

\subsection{The Metric Perturbation}
As it is well known in the literature of preheating, see e.g. Refs.~\citep{Kodama:1996jh,Bassett:1999mt,Giblin:2019nuv,Eggemeier:2023nyu,Chung:1999ve,Pearce:2017bdc},
resonant processes also affect the metric perturbation. The current
model is not an exception. In order to estimate these effects and
to compute the final spectrum of the primordial curvature perturbation
we employ semi-analytic computations. To that goal several simplifications
are made. First of all, we will only solve linearised equations. It
is likely that such an approximation provide sufficiently accurate
results. In contrast to the preheating scenarios, perturbations during
inflation must remain linear. This also justifies using the Hartree
approximation to estimate the effects of non-linear terms. Due to
the smallness of perturbations, we would expect non-linear $k$-mode
interactions of the metric perturbation to not change the picture
significantly.

At the linear level, we perform the computations in the Newtonian
and flat gauges. The two gauges are used in order to check the
consistency of our numerical code. We present Newtonian gauge equations
in this section and analogous expressions in the flat gauge in Appendix ~\ref{sec:flat_gauge}. The line element in the former takes the form
\begin{eqnarray}
\mathrm{d}s^{2} & = & -\left(1+2\Phi\right)\mathrm{d}t^{2}+a^{2}\left(t\right)\left(1+2\Psi\right)\delta_{ij}\mathrm{d}x^{i}\mathrm{d}x^{j}\,.\label{ds2-newt}
\end{eqnarray}
Since this is a two scalar field model in General Relativity, the
anisotropic stress vanishes and the two metric perturbation variables
are related by $\Psi=-\Phi$. Therefore we can drop $\Psi$ in favour
of $\Phi$. 

Scalar fields $\phi$ and $\chi$ are also perturbed such that
\begin{eqnarray}
\phi\left(\boldsymbol{x},t\right) & = & \bar{\phi}\left(t\right)+\delta\phi\left(\boldsymbol{x},t\right)
\end{eqnarray}
and
\begin{eqnarray}
\chi\left(\boldsymbol{x},t\right) & = & \bar{\chi}\left(t\right)+\delta\chi\left(\boldsymbol{x},t\right)\,.
\end{eqnarray}
In the case of the $\phi$ field we have $\bar{\phi}\gg\delta\phi$,
therefore the separation into the homogeneous value $\bar{\phi}$
and the perturbation $\delta\phi$ is unambiguous. In regards to the
$\chi$ field, an analogous separation is more subtle. Initially $\chi$
is heavy and its VEV vanishes. Hence, we define $\bar{\chi}$ by
\begin{equation}
\bar{\chi}\equiv\sqrt{\left\langle \chi^{2}\right\rangle }\,,\label{chi-hom}
\end{equation}
where $\left\langle \chi^{2}\right\rangle $ is given in Eq.~\eqref{chi2}
and we take $\delta\chi$ to be of the same perturbation order as
$\delta\phi$. Finally, because we have no use of the full fields
$\phi\left(\boldsymbol{x},t\right)$ and $\chi\left(\boldsymbol{x},t\right)$,
we will drop the overbars from the homogeneous fields and denote them
just by $\phi$ and $\chi$ in the remaining part of the text.

The homogeneous components follow the equations
\begin{align}
\ddot{\phi}+3H\dot{\phi}+V_{,\phi} & =0\,,\label{eom-phi}\\
\ddot{\chi}+3H\dot{\chi}+V_{,\chi} & =0\,,\label{eom-chi}
\end{align}
where the Hubble parameter is given by
\begin{equation}
3H^{2}=\frac{1}{2}\dot{\phi}^{2}+\frac{1}{2}\dot{\chi}^{2}+V\label{H2}
\end{equation}
and $V$ denotes the full potential
\begin{equation}
V\equiv V\left(\phi\right)+V\left(\chi\right)+V_{\mathrm{int}}\left(\phi,\chi\right)\,.\label{V}
\end{equation}

In regards to perturbations, the full system of equations in the Newtonian
gauge is given by 
\begin{align}
\delta\ddot{\phi}_{k}+3H\delta\dot{\phi}_{k}+\left(\frac{k^{2}}{a^{2}}+V_{,\phi\phi}\right)\delta\phi_{k} & =2\left(2\dot{\phi}\dot{\Phi}_{k}-V_{,\phi}\Phi_{k}\right)-V_{,\phi\chi}\delta\chi_{k}\,,\label{dphi-eqn}\\
\delta\ddot{\chi}_{k}+3H\delta\dot{\chi}_{k}+\left(\frac{k^{2}}{a^{2}}+V_{,\chi\chi}\right)\delta\chi_{k} & =2\left(2\dot{\chi}\dot{\Phi}_{k}-V_{,\chi}\Phi_{k}\right)-V_{,\phi\chi}\delta\phi_{k}\,,\label{dchi-eqn}\\
\dot{\Phi}_{k}+H\Phi_{k} & =\frac{1}{2}\left(\dot{\phi}\delta\phi_{k}+\dot{\chi}\delta\chi_{k}\right)\,,\label{dPhi-eqn}
\end{align}
where $\delta\phi_{k}$, $\delta\chi_{k}$ and $\Phi_{k}$ represent
the Fourier modes of perturbation variables $\delta\phi$, $\delta\chi$
and $\Phi$ respectively. We use Eq.~(\ref{dPhi-eqn}) in the integral
form to inspect numerical solutions. In this form the equation can
be written as
\begin{equation}
\Phi_{k}=\frac{1}{2a}\int a\left(\dot{\phi}\delta\phi_{k}+\dot{\chi}\delta\chi_{k}\right)\mathrm{d}t\,.
\end{equation}
In addition, the perturbed Einstein equation results in a constraint
equation 
\begin{equation}
\left(\dot{\phi}^{2}+\dot{\chi}^{2}-2\frac{k^{2}}{a^{2}}\right)\Phi_{k}=\dot{\phi}\delta\dot{\phi}_{k}+\dot{\chi}\delta\dot{\chi}_{k}-\ddot{\phi}\delta\phi_{k}-\ddot{\chi}\delta\chi_{k}\,.
\end{equation}

Similarly to Eq.~(\ref{dchi-eom-nog}), using the Hartree approximation
we replace $\chi^{2}$ with $\left\langle \chi^{2}\right\rangle $
whenever such a term appears in Eqs.~(\ref{dphi-eqn})--(\ref{dPhi-eqn}).
For example, 
\begin{equation}
V_{,\chi\chi}=g^{2}\left(\phi-\phi_{\c}\right)^{2}-m^{2}+12\lambda\left\langle \chi^{2}\right\rangle \,.
\end{equation}

Eq.~\eqref{dPhi-eqn} makes it clear that $\Phi_{k}$ is directly
sourced by the trapping field perturbation $\delta\chi_{k}$. Hence,
if $\delta\chi_{k}$ is resonantly amplified, one expects that it
also amplifies the metric perturbation. As we will see bellow, this
is exactly what happens during the resonance.

The ultimate goal of solving these equations is to compute the primordial
curvature perturbation $\zeta$. In terms of the Newtonian metric
perturbation it is given by 
\begin{equation}
\zeta_{k}=\Phi_{k}+2H\frac{\dot{\Phi}_{k}+H\Phi_{k}\left(1+\frac{1}{3}\frac{k^{2}}{a^{2}H^{2}}\right)}{\dot{\phi}^{2}+\dot{\chi}^{2}}\,.\label{z-newt}
\end{equation}
The power spectrum of $\zeta_{k}$ is then computed using
\begin{equation}
\mathcal{P}_{\zeta}\left(k\right)=\frac{k^{3}}{2\pi^{2}}\left|\zeta_{k}\right|^{2}\,.\label{Pz-def}
\end{equation}

Since we consider the two field model, inevitably the isocurvature
perturbation is also generated at some level. Such a perturbation
can be computed using the following expression \citep{Gordon:2000hv}
\begin{equation}
\mathcal{S}_{k}=\frac{2}{3}\frac{\delta V\left[3H\left(\dot{\phi}^{2}+\dot{\chi}^{2}\right)+\dot{V}\right]+\left[\dot{\phi}\delta\dot{\phi}_{k}+\dot{\chi}\delta\dot{\chi}_{k}-\Phi_{k}\left(\dot{\phi}^{2}+\dot{\chi}^{2}\right)\right]\dot{V}}{\left(\dot{\phi}^{2}+\dot{\chi}^{2}\right)\left[3H\left(\dot{\phi}^{2}+\dot{\chi}^{2}\right)+2\dot{V}\right]}\,,
\end{equation}
where $\delta V = V_{,\phi}\delta \phi + V_{,\chi}\delta\chi$. Similarly to Eq.~(\ref{Pz-def}) we define the spectrum of the isocurvature
perturbation to be
\begin{equation}
\mathcal{P}_{\mathcal{S}}\left(k\right)=\frac{k^{3}}{2\pi^{2}}\left|\mathcal{S}_{k}\right|^{2}\,.
\label{PS-def}
\end{equation}

Because the $\chi$ field is heavy when the pivot scale exits the horizon,
$\mathcal{S}_{k}$ is negligible on those scales (see Fig. \ref{fig:numP}
for an example). 
On smaller length scales, for modes exiting the horizon during the trapping phase, this is no longer true. 
As it is well known (see for example Ref.~\citep{Gordon:2000hv}) a non-zero isocurvature perturbation
can source the curvature one, even on superhorizon scales. 
But this depends on the reheating scenario and other factors. 
Since we assume prompt reheating for the purpose of this work, we do not include the
contribution from $\mathcal{S}_{k}$ to $\zeta_{k}$ during the post inflationary evolution. 
The study of these effects and their consequences for the mass distribution of PBHs is left for future work.

Furthermore, as pointed out in ~\cite{Murata:2025onc}, the stochastic effect may become dominant in hybrid inflation with a waterfall under certain conditions. Therefore, we compare the classical effects, $m_\text{Pl}^2V_{,\phi}/V$ and $m_\text{Pl}^2V_{,\chi}/V$, in the Friedmann equations with the quantum effect, $H/2\pi$~\cite{Martin:2011ib}. We calculated the ratios defined as $\Delta_\phi \equiv H V/2\pi m_\text{Pl}^2 V_{,\phi}$ and $\Delta_\chi \equiv H V/2\pi m_\text{Pl}^2 V_{,\chi}$. In the case of our hybrid inflation potential, both $\Delta_\phi\sim10^{-10}$ and $\Delta_\chi\sim10^{-15}$ were found to be small, indicating that the stochastic effect is negligible and a classical treatment is justified.

\section{The Second Stage and The Total Duration of Inflation\label{duration}}
One of the traditional issues related to RMI is its large spectral
running \citep{Alabidi:2009bk,Kohri:2018qtx}. Usually it takes only
a few tens of e-folds of inflation before perturbations become non-linear.
Such a short inflation is not sufficient if it is to solve the horizon
and flatness problems. The duration of inflation can be enlarged if
the $V\left(\chi\right)$ part of the potential is flat enough, so
that the waterfall phase can provide the missing number of e-folds.
This is in contrast to the standard picture of hybrid inflation \citep{Linde:1993cn,Copeland:1994vg},
where the waterfall phase is assumed to be completed within less than
an e-fold.

For this purpose we consider a hilltop type potential in Eq.~(\ref{inf2V}),
which consists of two free parameters $m$ and $\lambda$. Only these
two terms are assumed to be significant during the waterfall. Higher
order terms can be added to stabilise the potential, but they are
taken to be inconsequential for the dynamics of inflation.

\begin{figure}
\begin{centering}
\includegraphics[scale=0.5]{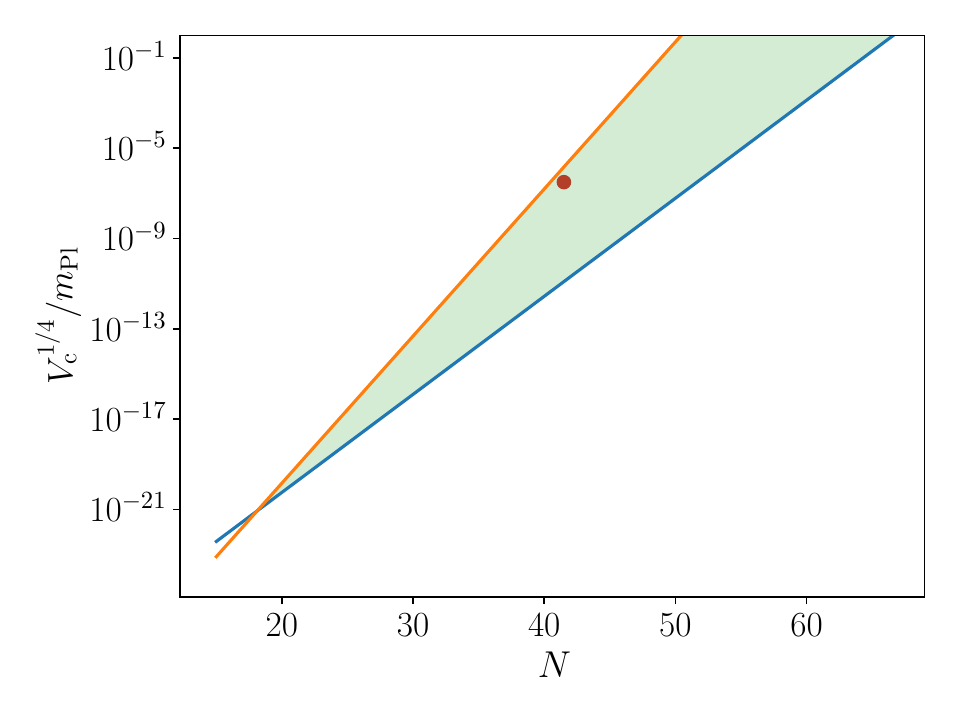}
\par\end{centering}
    \caption{\label{fig:infE}The allowed range of the energy scale of inflation 
    (green shaded region). Orange and blue lines represent bounds in 
    Eqs.~\eqref{V-lower} and \eqref{V-upper} respectively. 
    The red dot marks the value of the model specified in 
    section~\ref{subsec:numerical}.
    }
\end{figure}

To estimate the minimum number of e-folds of inflation that is required
to solve the flatness and horizon problems of HBB we assume
prompt reheating at the end of inflation. In this approximation we
can write \citep{Lyth:2009zz}
\begin{equation}
N\simeq56-\frac{2}{3}\ln\frac{10^{16}\text{ GeV}}{\rho_{*}^{1/4}}-\frac{1}{3}\ln\frac{10^{9}\text{ GeV}}{T_{\reh}}\,,
\end{equation}
where $N$ is the number of e-folds defined by $N\equiv\ln a/a_{0}$,
$\rho_{*}^{1/4}$ is the energy scale of inflation when the pivot
scale leaves the horizon in units of GeV and $T_{\reh}$ is the temperature
at reheating, also in GeV. We can invert this expression and write
\begin{equation}
\frac{T_{\reh}}{\text{ MeV}}\simeq1.7\times10^{7}\:\mathrm{e}^{3\left(N-56\right)}\frac{\mpl^{2}}{\sqrt{V_{\mathrm{c}}}}\,,\label{Treh}
\end{equation}
where $V_{\mathrm{c}}$ is defined in Eq.~(\ref{rmiV}) and we used
the fact that $U\left(\phi\right)\ll1$. There are (at least) two
conditions that this equation must satisfy. First, the reheating temperature
must be larger than the temperature of the Big Bang Nucleosynthesis,
which is $T_{\mathrm{BBN}}\sim1\text{ MeV}$ \citep{Kawasaki:1999na,Hannestad:2004px,deSalas:2015glj}.
It follows from the above equation that the upper bound on the energy
scale of inflation must be
\begin{equation}
    V_{\mathrm{c}}<3\times10^{14}\cdot\mathrm{e}^{6\left(N-56\right)}\mpl^{4}\,.\label{V-lower}
\end{equation}

On the other hand, for a given duration and the energy scale of inflation,
one must make sure that blindly applying Eq.~(\ref{Treh}) does not
lead to the energy density $\rho_{\reh}$ at reheating to become larger
than the energy scale at the end of inflation. For this estimate it
will be sufficient to assume constant energy density during inflation
and use the relation of the thermalised radiation
\begin{equation}
\rho_{\reh}=\frac{\pi^{2}g_{*}}{30}T_{\reh}^{4}\,,
\end{equation}
where $g_{*}$ is the effective number of relativistic degrees of freedom. 
At temperatures $T>100$~GeV this number is $g_* = \mathcal{O}\left(100\right)$.
Thus, we find from Eq.~(\ref{Treh}) that the condition
$V_{\mathrm{\mathrm{end}}}>\rho_{\reh}$, where $V_{\mathrm{end}}$
is the energy scale at the end of inflation, leads to the inequality
\begin{equation}
V_{\mathrm{c}}>10^{-19}\left(\frac{\pi^{2}g_{*}}{30}\right)^{\frac{1}{3}}\mathrm{e}^{4\left(N-56\right)}\mpl^{4}\,,\label{V-upper}
\end{equation}
where we took $\left(V_{\mathrm{c}}/V_{\mathrm{end}}\right)^{\frac{1}{3}}\sim1$.

Putting Eqs.~(\ref{V-lower}) and (\ref{V-upper}) together and taking
$g_{*}=150$ for concreteness and a rough estimate, we find
\begin{equation}
5\times10^{-19}\mathrm{e}^{4\left(N-56\right)}<\frac{V_{\mathrm{c}}}{\mpl^{4}}<3\times10^{14}\cdot\mathrm{e}^{6\left(N-56\right)}\,.
\end{equation}
This bound is easier to appreciate looking at Fig.~\ref{fig:infE}.
When searching for a viable parameter space of this model,
the above condition, together with Eqs.~\eqref{ns-const}--\eqref{r-const},  needs to be satisfied.

The effects of radiative corrections to the dynamics of the waterfall
field \citep{Gong:2022tfu} are ignored in this study. We don't expect
such corrections to change the picture qualitatively. But their effect
on the space of allowed parameter values should certainly be studied,
which we plan to do in the future.

\section{The Primordial Curvature Perturbation\label{subsec:numerical}}
\begin{figure}
\begin{centering}
\includegraphics[scale=0.5]{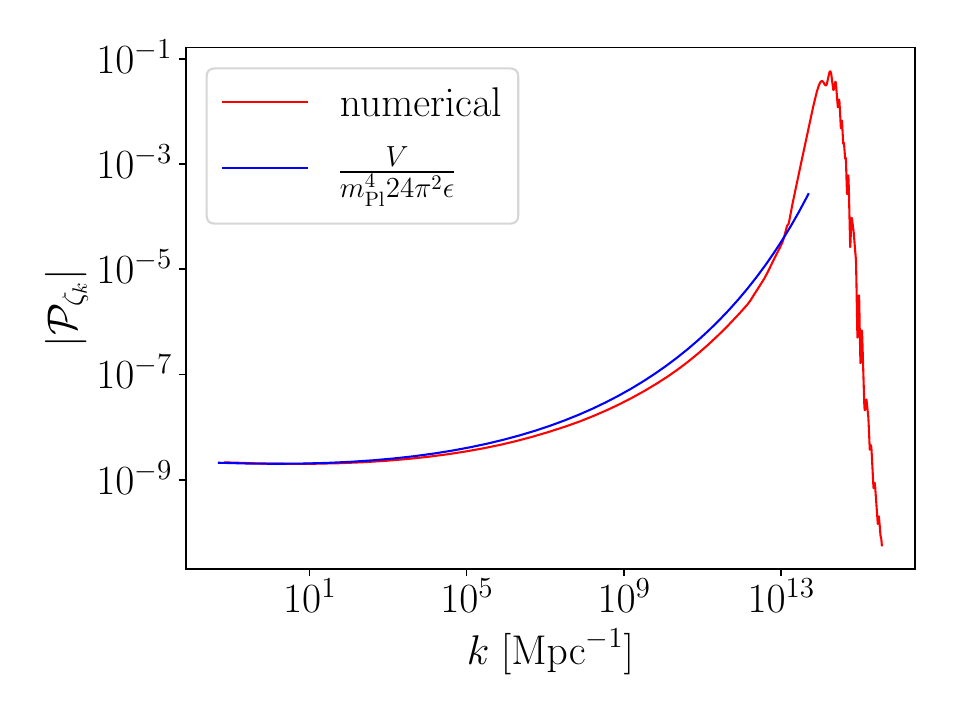} 
~ \includegraphics[scale=0.5]{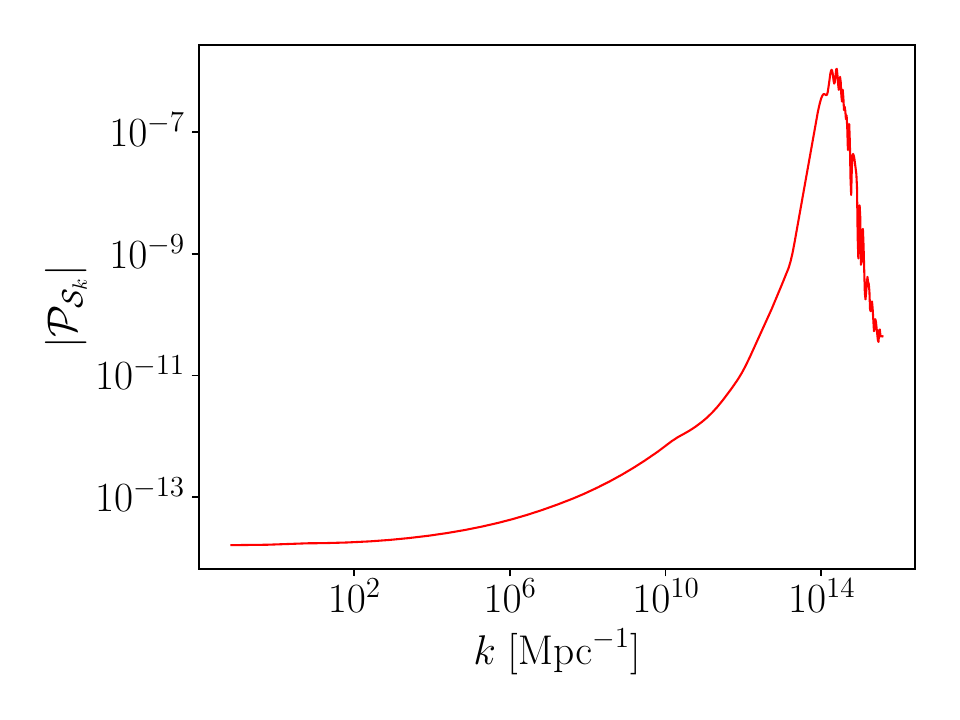}
\par\end{centering}
\caption{\label{fig:numP}
    Left panel: the primordial curvature perturbation spectrum $\mathcal{P}_{\zeta}\left(k\right)$ (see Eq.~\eqref{Pz-def}). 
    The red curve indicates numerical results. 
    For comparison we also show (the blue curve) the spectrum computed using the slow-roll approximation (see Eq.~\eqref{Pslow}).
    This curve terminates at the scales that exit the horizon when $\phi=\phi_{\c}$.
    Right panel: the spectrum of the isocurvature perturbation $\mathcal{P}_{\mathcal{S}}\left(k\right)$ (see Eq.~\eqref{PS-def})
}
\end{figure}

To find models that are compatible with observations and provide large
enough primordial perturbation on small scales, we perform numerical
simulations. To do that, we first estimate the spectrum
using slow-roll approximation for all models in the allowed regions
shown in Fig.~\ref{fig:inf-constr}. This narrows down the set of
models which are likely to produce the correct amplitude of the spectrum
at the required scales.

We next perform numerical simulations of the exact linear equations
Eqs.~(\ref{eom-phi})--(\ref{dPhi-eqn}) applied to this narrowed down set of models.
Our goal is to find models that give the spectrum with a sharp peak of amplitude 
$\mathcal{P}_{\zeta}\left(k_{\max}\right)\simeq10^{-1.5}$ at around 35 e-folds after the pivot scale exits the horizon. 
Such values are likely to lead to the correct mass distribution of PBHs, as discussed in sec.~\ref{sec:pbh}. 

The parameters of one such model, which we continue using for 
the rest of the paper, are $\alpha = 0.005$, $A = 4.849$, $B = 5.859$. 
We found that for this model CMB constraints in Eqs.~\eqref{ns-const}--\eqref{r-const} are
best satisfied when the inflaton field value is $\phi_* = 3.89\times10^{-9}\,\mpl$.
Consequently, this leads to the energy scale of inflation $V_\mathrm{c}^{1/4} = 10^{-6.5}\,\mpl$. 
$N=35$ e-folds later the inflaton reaches $\phi_{\c} = 6.02\times10^{-13}\,\mpl$. 
At this moment the trapping field is resonantly excited and rendered unstable. 
We ran a number of simulations to search for parameter values of the waterfall potential that give the right
value of $\mathcal{P}_{\zeta}$ and guarantee a long enough waterfall phase. 
One such possible model resulted in the trapping field mass $m=3.3\times10^{-13}\mpl$ and $g^2=0.81$. 
The quartic self-coupling strength $\lambda$ is chosen such that the vacuum energy vanishes, i.e. 
$\lambda = 5/24\cdot m^4/V\left(\phi_c\right) = 1.14\times10^{-24}$. 

The duration of inflation from the moment the pivot scale exits the horizon to the end of inflation is $N=41.5$ e-folds in this model.
It is somewhat shorter than the conventional range from 50 to 60 e-folds. 
Nevertheless this value is sufficient to solve HBB problems, as detailed in section~\ref{duration}. 
Indeed, the discussed model falls within the green region of Fig.~\ref{fig:infE}.

The numerically computed spectrum is shown in the left panel of Fig.~\ref{fig:numP}.
In that plot we also provide the spectrum (the blue curve) computed using
slow-roll approximation \citep{Lyth:2009zz}
\begin{eqnarray}
    \mathcal{P}_{\zeta}\left(k\right) & = & \frac{1}{24\pi^{2}\mpl^{4}}\left.\frac{V}{\epsilon}\right|_{k}\,,\label{Pslow}
\end{eqnarray}
where $\epsilon$ is defined in Eq.~\eqref{epsilon-def} and the index
`$k$' indicates that $\epsilon$ and $V$ values must be evaluated at the horizon
crossing. 
As one expects, this expression provides a good approximation of the spectrum for small $k$, but it starts deviating
from the more accurate, numerically computed spectrum once the resonant production of $\chi$
particles commences. 

In the right panel of Fig.~\ref{fig:numP} we also show the spectrum of the isocurvature perturbation. 
As one can see, it is negligible on the CMB scales (small $k$ values), which is required in order
to satisfy the tight bounds on this mode from Planck constraints \citep{Planck:2018jri}. 

In addition to the Newton gauge expressions, we also perform the same
simulations for perturbations in the flat gauge (see Appendix ~\ref{sec:flat_gauge}),
which provides a check of our computations. 
The results of the latter are not shown, because they are virtually indistinguishable from the
Newtonian gauge ones.

\section{Mass Distribution of Primordial Black Holes\label{sec:pbh}}
In this section we calculate the mass function (i.e., the mass
distribution) of PBHs as predicted by the current model.
Roughly speaking we need the curvature perturbation to be of order
$\mathcal{P}_{\zeta}\left(k\right) \sim {\cal O}(10^{-1.5})$, so that during 
radiation domination PBHs are produced via gravitational collapse. 
Then, the relation
between the mass of PBHs and the wave number $k$ can be written by
\begin{eqnarray}
  \label{eq:mPBH-k-relation}
  m_{\mathrm{PBH}} \sim 10^{20} {\rm g}\left(\frac{k_*}{ 10^{14} \mathrm{Mpc}^{-1}}\right)^{-2}.
\end{eqnarray}
This corresponds to $N\sim 35$ e-folds after the pivot scale exits the horizon.

Having the spectrum of the primordial curvature perturbation $\mathcal{P}_\zeta$ 
(see Fig.~\ref{fig:numP}) we can compute the abundance of PBHs following, for example, Ref.~\cite{Kohri:2018qtx}. 

First, let us define the fraction of the energy density of PBHs relative to that 
of Cold Dark Matter (CDM) evaluated at present time
\begin{equation}\label{eq:fdef}
f_\text{PBH} \equiv \frac{\rho_\text{PBH}}{\rho_\text{CDM}},
\end{equation}
where $\rho_\text{PBH}$ and $\rho_\text{CDM}$ denote energy densities of PBHs and CDM respectively. 
Then the mass function per logarithmic bin in mass
$d f_\text{PBH}(m_{\mathrm{PBH}})/d \ln({m_{\mathrm{PBH}})} \sim
f_\text{PBH}(m_{\mathrm{PBH}})$ can be expressed as
\begin{equation}\label{eq:fm}
f_{\text{PBH}}(m_{\mathrm{PBH}})= \frac{\Omega_{\text{m}}}{\Omega_{\text{CDM}}} \left[ \frac{g_{*}(T)}{g_{*}(T_{\text{eq}})} \frac{g_{*,s}(T_{\text{eq}})}{g_{*,s}(T)} \frac{T(m_{\mathrm{PBH}})}{T_{\text{eq}}} \gamma \beta(m_{\mathrm{PBH}}) \right],
\end{equation}
where we used the fraction $\beta$ of the energy density of PBHs
$\rho_\text{PBH}$ relative to the total energy density
$\rho_\text{tot}$ at the formation epoch, 
$\beta \equiv {\rho_\text{PBH}}/{\rho_\text{tot}}$. In the above expression,
$\Omega_{\text{m}}$ and $\Omega_{\text{CDM}}$ denote cosmological density
parameters of matter and CDM respectively. For these parameters we adopt the
values reported by the Planck team in Ref.~\cite{Planck:2018jri}. Also,
$g_{*}$, $g_{*,s}$ denote the number of relativistic degrees of freedom
that contribute to the energy and entropy densities respectively. We use their concrete
time-dependent values as reported in Ref.~\cite{Saikawa:2020swg}. Temperatures 
$T(m_{\mathrm{PBH}})$ and $T_\text{eq}$ are evaluated 
at the formation of PBHs and at the matter-radiation equality respectively, while
constant $\gamma$ denotes the ratio between the mass of the PBHs  
$m_{\mathrm{PBH}}$ and the horizon mass
$M_\text{H} = \frac{4 \pi}{3}\frac{\rho}{H^3}$ given by
\begin{equation}\label{eq:gamma}
m_{\mathrm{PBH}} = \gamma M_\text{H},
\end{equation}
where the energy density $\rho$ is computed using the Friedmann equation, 
$\rho = 3 m^2_\text{Pl} H^2$, and $H$ is the Hubble parameter evaluated at the
time of PBH formation. 
According to a simple analytic formula, the value of $\gamma$ is estimated to be
$\gamma=(1/\sqrt{3})^3 \sim 0.2$~\cite{1975ApJ...201....1C}. 

Assuming that the density perturbation follows Gaussian distribution, we can
compute $\beta$ using the Press Schechter theory~\cite{Press:1973iz}, which gives
\begin{equation}
\label{eq:bm}
\beta (m_{\mathrm{PBH}}) = \int_{\delta_\text{c}}^{\infty} \frac{\text{d}\delta}{\sqrt{2\pi}\sigma(m_{\mathrm{PBH}})} \exp \left[ \frac{-\delta^2}{2\sigma^2(m_{\mathrm{PBH}})} \right]= \frac{1}{2} \text{Erfc}\left[ \frac{\delta_{\text{c}}}{\sqrt{2}\sigma(m_{\mathrm{PBH}})} \right].
\end{equation}
As it is clear from the above, $\beta$ is a function of the PBH mass, similarly to
Eq.~(\ref{eq:fm}). Here $\mathrm{Erfc}$ denotes the complementary error
function. The threshold $\delta_{\text{c}}$ represents the critical
value for PBH formation. The analytical expression for this quantity was computed
in Ref.~\cite{Harada:2013epa}. In this work, we adopt the value
$\delta_{\text{c}} = 0.45$~\cite{Musco:2004ak}. The density perturbation that exceeds this
threshold value undergoes gravitational collapse when it re-enters the Hubble
horizon, leading to the formation of PBHs. 

Furthermore, applying the asymptotic expansion of the complementary
error function in Eq.~(\ref{eq:bm}), we obtain
\begin{equation}\label{eq:bmasy}
\beta (\sigma)\simeq \frac{\sigma(m_{\mathrm{PBH}})}{\sqrt{2\pi}\delta_c} \exp \left[\frac{-\delta^2_c}{2\sigma^2(m_{\mathrm{PBH}})}\right],
\end{equation}
where $\sigma$ is the coarse-grained density perturbation given by
\begin{equation}\label{eq:sig}
\sigma^2(k)=\int_{-\infty}^{\infty} \mathrm{d} \ln q W^2\left(\frac{q}{k}\right) \frac{4\left(1+w_{\mathrm{eos}}\right)^2}{\left(5+3 w_{\mathrm{eos}}\right)^2}\left(\frac{q}{k}\right)^4 \mathcal{P}_\zeta(q),
\end{equation}
and $w_{\mathrm{eos}}$ denotes the equation of state parameter,
which is defined by $w_{\mathrm{eos}} = p/\rho$, while $W$ denotes the window
function, which is taken to be $W(k) = \exp(-k^2/2)$.

\begin{figure}
\begin{centering}
    \includegraphics[width=10cm]{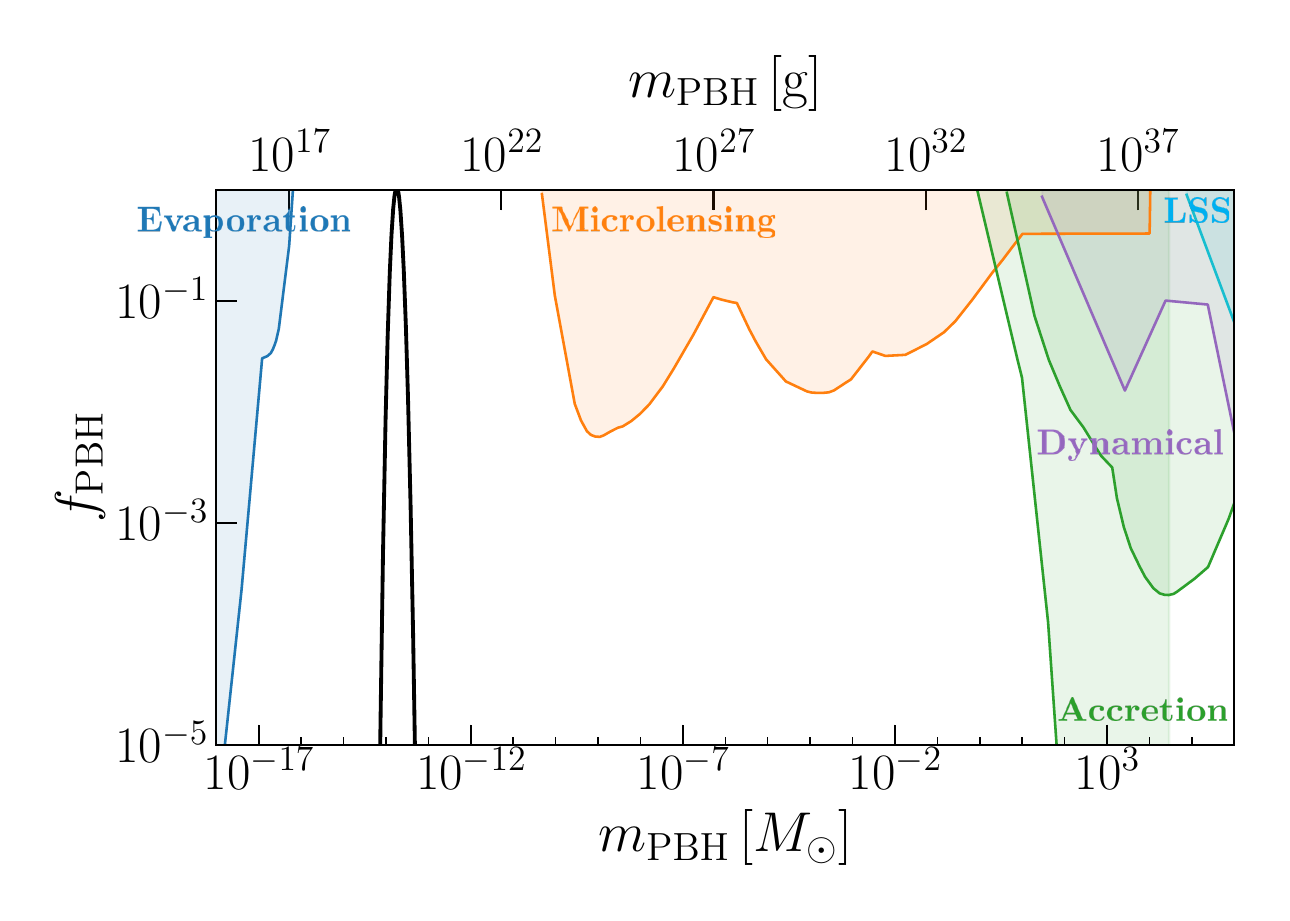}
\par\end{centering}
\caption{The mass distribution of PBHs (black curve) calculated by the 
spectrum given in Fig.~\ref{fig:numP} as a function of the PBH mass. The 
vertical axis indicates the energy fraction of
PBHs to the CDM ($f_{\text{PBH}}$). The colored curves represent
observational upper bounds on
$f_{\text{PBH}}$~\cite{Carr:2020gox,Carr:2009jm}. The blue curve
corresponds to constraints from the evaporation of the PBHs,
including the extragalactic $\gamma$-ray background
(EGB)~\cite{Carr:2020gox,Carr:2009jm}, the Voyager positron flux
(V)~\cite{Boudaud:2018hqb}, and annihilation-line radiation from the
Galactic Centre (GC)~\cite{Carr:2016hva}. The orange curve shows
constraints from gravitational lensing, including those from 
supernovae (SN)~\cite{Graham:2015apa}, the M31 stars observed by
Subaru/HSC~\cite{Smyth:2019whb}, the Magellanic Clouds by EROS and
MACHO (EM)~\cite{EROS-2:2006ryy, Macho:2000nvd}, and the Galactic
bulge by the OGLE (O)~\cite{Niikura:2019kqi}. The green curve shows
the constraints from accretion, including X-ray binaries
(XB)~\cite{Inoue:2017csr} and the spectral distortions of the CMB
measured by Planck (PA)~\cite{Tashiro:2008sf}. The purple curve
represents the dynamical constraints, including those from wide binaries
(WB)~\cite{Monroy_Rodr_guez_2014}, star clusters in Eridanus II
(E)~\cite{Zoutendijk_2020}, halo dynamical-friction
(DF)~\cite{Carr:2018rid}, galaxy tidal-distortions
(G)~\cite{1999ApJ...516..195C}, heating of stars in the Galactic
disk (DH)~\cite{Carr:2018rid}, and the CMB dipole (CMB). The cyan curve indicates constraints from large-scale structure
formation~\cite{Carr:2018rid, Murgia:2019duy}.}
\label{fig:fpbh}
\end{figure}

Putting all together, the PBH abundance 
reaches the value $ f_{\text{PBH}} \sim \mathcal{O}(1)$ in the current model.
Moreover, the mass of the PBHs
peaks at $m_{\text{PBH}} \sim 10^{19.5}$~g which is within the
asteroid-mass range
($10^{17}\text{g} \lesssim m_{\text{PBH}} \lesssim 10^{23}\text{g}$)
where the observational upper limits (see the colored curves in 
Fig.~\ref{fig:fpbh}) still allow for the possibility of PBHs being 100$\%$ of 
CDM.

\section{Induced Gravitational Waves}\label{sec:sigw}
\begin{figure}
\begin{centering}
\includegraphics[width=10cm]{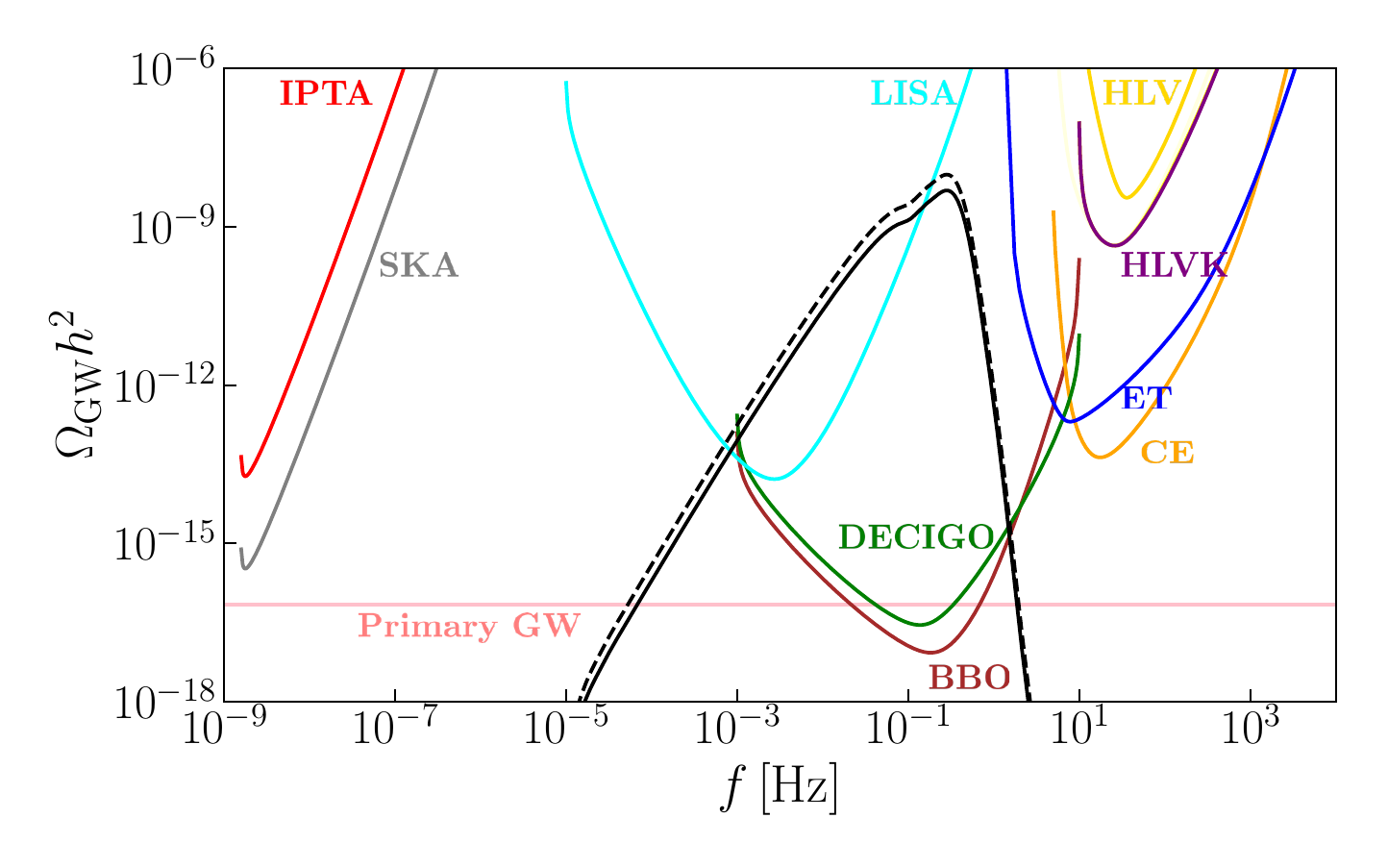}
\par\end{centering}
\caption{Energy density of gravitational waves $\Omega_{\text{GW}}h^2$
    as a function of the frequency $f$ in units of Hz, where $h$ is the
    dimensionless Hubble constant. The black dashed curve represents the 
    spectrum of SIGWs, the black solid curve shows GWs with the dissipation, 
    and the pink curve marks the approximate amplitude of primary GWs.
    Other colored curves show the sensitivities of various gravitational wave 
    observatories~\cite{Schmitz:2020syl}. They are
    the International Pulsar Timing Array (IPTA)  (red)~\cite{Hobbs_2010,
    Manchester_2013, Verbiest_2016, Hazboun:2018wpv}, the Square
    Kilometre Array (SKA)  (grey)~\cite{Carilli:2004nx, Janssen:2014dka,
    Weltman:2018zrl}, the Laser Interferometer Space Antenna
    (LISA)  (cyan)~\cite{amaroseoane2017laserinterferometerspaceantenna,
    Baker:2019nia}, the Deci-Hertz Interferometer
    Gravitational-Wave Observatory (DECIGO)  (green)~\cite{Seto:2001qf,
    Kawamura:2006up, Yagi:2011wg, Isoyama:2018rjb}, 
    the Big-Bang Observer (BBO)  (brown)~\cite{Isoyama:2018rjb,
    Crowder:2005nr, Corbin:2005ny, Harry:2006fi},  the Hanford-Livingston-Virgo (HLV)  (yellow)~\cite{HLV1, HLV2, HLV3}, the Hanford-Livingston-Virgo-KAGRA
    (HLVK)  (purple)~\cite{HLVK1, HLVK2}, the Einstein Telescope
    (ET)  (blue)~\cite{Punturo:2010zz, Hild:2010id, Sathyaprakash:2012jk,
    ET:2019dnz}, and the Cosmic Explorer
    (CE)  (orange)~\cite{LIGOScientific:2016wof, Reitze:2019iox}.
    }
\label{fig:sigw}
\end{figure}

The large amplitude of the curvature perturbation $\mathcal{P}_\zeta$ on small
scales, which is responsible for the formation of PBHs, is also
responsible for the (stochastic) induced gravitational waves (SIGWs) that are 
generated via nonlinear second-order effects. 
In this section, we calculate the spectrum of such GWs
that are produced during the radiation-dominated epoch.
For the computation we follow the method detailed in 
Refs.~\cite{Kohri:2018awv,Sasaki:2018dmp,Terada:2025cto}
and adapt it to the current model. 
The detailed calculation is presented in Appendix~\ref{app:sigw}, where the 
present day spectrum is found to be
\begin{equation}\label{eq:sigw}
\Omega_{\mathrm{GW}}(\eta, k)=\frac{\rho_{\mathrm{GW}}(\eta,k)}{\rho_{\mathrm{tot}}(\eta)}=\frac{1}{24}\left(\frac{k}{a(\eta) H(\eta)}\right)^{2} \overline{\mathcal{P}_{T}(\eta, k)}.
\end{equation}
In this expression $\eta$ denotes the conformal time, and the wave number $k$ is related
to the frequency $f$ by $k = 2\pi f$. The quantity
$\rho_{\mathrm{GW}}(\eta,k)$ denotes the energy density of
GWs per logarithmic interval of the wave number, and
the overline $\overline{\mathcal{P}_T}$ indicates the oscillation
average of the tensor perturbation power spectrum
$\mathcal{P}_T(\eta, k)$. This quantity is defined as
\begin{equation}\label{eq:ph}
\mathcal{P}_{T}(\eta, k)=4 \int_{0}^{\infty} \mathrm{d} v \int_{|1-v|}^{1+v} \mathrm{~d} u\left(\frac{4 v^{2}-\left(1+v^{2}-u^{2}\right)^{2}}{4 v u}\right)^{2} I^{2}(v, u, x) \mathcal{P}_{\zeta}(k v) \mathcal{P}_{\zeta}(k u),
\end{equation}
where $x$ is the dimensionless variable $x = k\eta$, while
$u = |{\bf k} - \tilde{\bf k}|/k$ and $v = \tilde{k}/k$ are
integration variables representing the momentum configuration. 
The function $I(v,u,x)$ is an oscillating function from the source
information.
From the expression for GWs, the spectrum includes the scalar perturbations through the source term of the tensor perturbation. The mean free path of weakly interacting light particles such as neutrinos damps the scalar field perturbations on small scales during the radiation-dominated epoch. Therefore, it also affects induced gravitational waves~\cite{Domenech:2025bvr, Jeong:2014gna, Yu:2025cqu}. Next, we present the kernel function for the case with dissipative effect and discuss the expression of the kernel function without dissipation in Appendix~\ref{app:sigw}, which is shown by the black dashed curve in Fig.~\ref{fig:sigw}. The effect on the oscillating function $I$ of the source term is given by
\begin{align}
I_j^{(0)} =&\, -\frac{1-c_s^2 \left(u^2+v^2\right)}{2 c_s^4 u^2 v^2} \Bigg(1 - \frac{1-c_s^2 \left(u^2+v^2\right)}{4 c_s^2 u v} \big[\,{\rm cei}[(1-c_s(u-v))]+{\rm cei}[(1+c_s(u-v))] \\
& -{\rm cei}[(1-c_s (u+v))]-{\rm cei}[(1+c_s (u+v))]\big] \Bigg), \nonumber\\
I_y^{(0)} =&\, \frac{\left(1-c_s^2 \left(u^2+v^2\right)\right)^2}{8 c_s^6 u^3 v^3} \big(\,{\rm Sei}[(1-c_s(u-v))]+{\rm Sei}[(1+c_s(u-v))] \nonumber \\
& -{\rm Sei}[(1-c_s (u+v))]-{\rm Sei}[(1+c_s (u+v))]\big)\,,
\end{align}
where the subscript $j$ and $y$ denote that the oscillating function is divided into terms of the spherical Bessel function of order zero. Furthermore, the superscript $(0)$ on $I$ denotes the term separated by the order of differentiation of $F=(k_D(\tau)/k_D(\tau_*))^{-2}$, which is normalized by the damping scale $k_D(\tau_*)$ at the pivot scale. The functions ${\rm cei}$ and ${\rm Sei}$ are given by
\begin{align}
{\rm cei}(y)&=\int_{0}^\infty \frac{dx}{x} e^{-{(u^2+v^2)\kappa_{D}^2}F[x/x_*]} \left[1-\cos (yx)\right],\\
{\rm Sei}(y)&=\int_{0}^\infty \frac{dx}{x} e^{-{(u^2+v^2)\kappa_{D}^2}F[x/x_*]} \sin (yx),\,
\end{align}
where $c_s^2=1/3$ is the sound speed, and $\kappa_D = k/k_D(\tau_*)$ is the dimensionless $k$ normalized by the damping scale at the pivot scale.

The result is shown in Fig.~\ref{fig:sigw}. The black solid curve in that figure
represents the spectrum of the induced GWs that are generated by scalar 
perturbations at second order in perturbation theory and are computed above,
while the dashed black curve is the same without disspiation effect. As pointed out in Ref.~\cite{Domenech:2025bvr, Jeong:2014gna, Yu:2025cqu}, these effect is also found to dissipate the peak and the low-frequency tail of GWs.
The pink line shows an approximate spectrum of primary GWs that are
generated by vacuum fluctuations during inflation. 
Other colored curves indicate sensitivity bounds of future planned GW 
observations (see the caption of Fig.~\ref{fig:sigw} for details). 
The figure shows that it becomes evident that our model predicts GWs which
fall within the detectability limits of LISA, DECIGO and BBO future observatories.

\section{Gravitational waves from merging  binary PBHs}\label{sec:gwmerg}
There is one more source of GWs. As binary PBHs merge they also induce a 
stochastic GW background \cite{Sasaki:2018dmp,Wang:2019kaf,Kohri:2024qpd}.
We compute the spectrum of such GWs in this section.

Details of the calculation are provided in Appendix~\ref{app:gwmerge}, where it
is shown that the spectrum obeys the following relation
\begin{equation}\label{eq:gwmerg}
\Omega_\text{GW}^\text{(merger)}(f) = \frac{f}{\rho_\text{c}} \int_0^{z_\text{sup}} \mathrm{d} z \, \frac{R(z)}{(1 + z) H(z)} \frac{\mathrm{d}E(f_\text{s})}{\mathrm{d}f_\text{s}}.
\end{equation}
In this expression $\rho_\text{c}$ denotes the critical energy density of the Universe.
$z$ is the redshift and $z_\text{sup}$ is the upper limit of
integration, which is computed as $z_\text{sup} = f_3/f - 1$, where $f_3$ is the
cutoff frequency of GW at the end of the ringdown phase of BH merger \cite{Sasaki:2018dmp}.

$R(z)$ represents the PBH merger rate and
${\mathrm{d}E(f_\text{s})}/{\mathrm{d}f_\text{s}}$ denotes the energy
spectrum of the gravitational wave emitted from the source. 
Detailed expressions of these functions are provided in 
Appendix~\ref{app:gwmerge}.

\begin{figure}
\begin{centering}
\includegraphics[width=10cm]{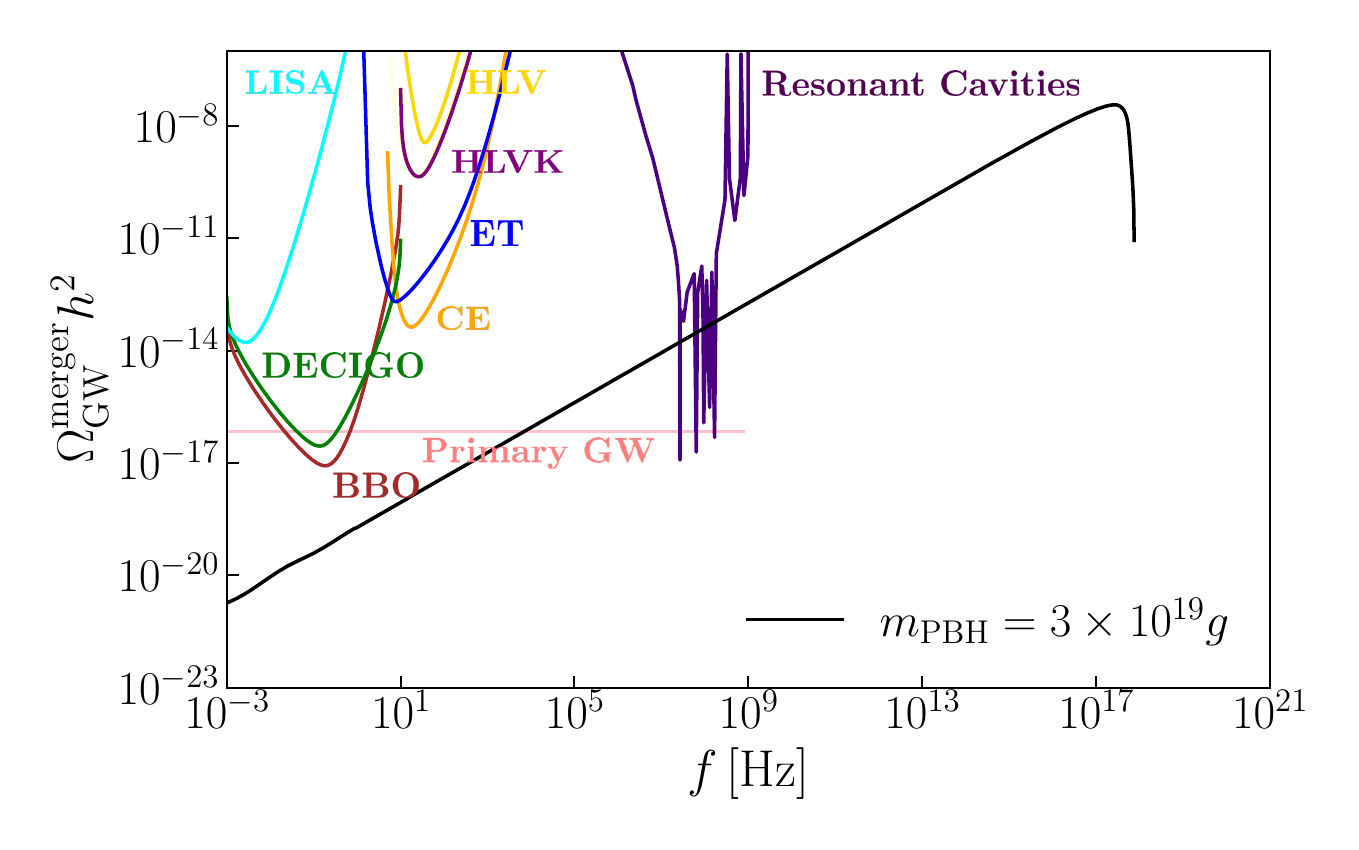}
\par\end{centering}
    \caption{The spectrum of GWs generated by PBH mergers (black curve). 
    Other coloured curves are the same as in Fig.~\ref{fig:sigw} 
    in addition to the resonant cavity experiment
    \cite{Berlin:2021txa,Herman:2022fau} (dark purple curve).}
\label{fig:gwmerg}
\end{figure}

As shown in Fig.~\ref{fig:fpbh}, the abundance of PBHs in our model exhibits a 
pronounced peak at $m_{\text{PBH}} \sim 3 \times 10^{19}\text{g}$ with
$f_{\text{PBH}} \sim \mathcal{O}(1)$. 
The GW spectrum that is produced by mergers of such PBHs is shown in
Fig.~\ref{fig:gwmerg}.
As can be seen in the figure, GW spectrum falls within the sensitivity 
region of resonant cavity detectors~\cite{Berlin:2021txa, Herman:2022fau}.

\section{Conclusions and Discussion}\label{sec:conc}
In this work we study a multi-field inflation scenario with a tachyonic trap in 
the context of supersymmetric running-mass-inflation models.
The scenario is reminiscent of hybrid inflation with some modifications. 
Initially the inflaton field rolls down the running-mass potential.
When the critical point $\phi_\mathrm{SBP}$ is reached, which we called the 
symmetry breaking point, the waterfall field is resonantly excited. 
Such excitations backreact onto
the motion of the inflaton field anchoring its value at $\phi_\mathrm{SBP}$.
The remaining number of e-folds of inflation are generated during the waterfall
phase. 
To make this phase long enough, the potential in the waterfall direction must be
sufficiently flat.

The proposed scenario enables us to model running-mass-inflation from the
time when observable scales exit the horizon to the end of inflation.
It also allows us to compute the spectrum of the curvature perturbation for the full
duration of inflation. 
We find that the spectrum exhibits a sharp peak, corresponding to the scales
which exit the horizon around SBP. The enhancement of the spectrum at these scales is due to the
shape of the running-mass potential. Additionally the resonance amplifies the
amplitude of the spectrum by several orders of magnitude more.

Such a large perturbation gives rise to PBHs with masses in the range  
$10^{17}$~g -- $10^{23}$~g, which are viable dark matter candidates. 
The same perturbation also sources induced gravitational
waves in the deci-Hz range, providing a natural signal for upcoming
space-based interferometers such as LISA, DECIGO, and BBO.
Furthermore, PBHs formed in this scenario can subsequently assemble
into binaries, whose mergers generate gravitational waves observable
both in resonant cavity experiments via the inverse Gertsenshtein
effect and in future space-based detectors.  Although we have not
explicitly addressed the effects of stochastic noise on the motion of the inflaton in this 
work, for the parameter ranges considered, we estimate such effects to remain 
subdominant, which justifies neglecting them.

\begin{acknowledgments}
This work was in part supported by JSPS KAKENHI Grants Nos. JP23KF0289, JP24K07027 (K.K.), MEXT KAKENHI Grants No. JP24H01825 (K.K.), and by the Spanish Research Agency (Agencia Estatal de Investigación) through national project CNS2022-13600, AEI/MCIU through grant PID2023-148162NB-C21 and ASFAE/2022/020 (A.S.).

\end{acknowledgments}

\appendix

\section{The Primordial Perturbation in the Flat Gauge\label{sec:flat_gauge}}
In the flat gauge the scalar part of the spatial curvature perturbation
vanishes. This allows us to write the perturbed line element as
\begin{eqnarray}
\mathrm{d}s^{2} & = & -\left(1+2\alpha\right)\mathrm{d}t^{2}-2a\beta_{,i}\mathrm{d}x^{i}\mathrm{d}t+a^{2}\left(\delta_{ij}+2\partial_{i}\partial_{j}\gamma\right)\mathrm{d}x^{i}\mathrm{d}x^{j}\,,
\end{eqnarray}
where $\alpha$ is the perturbation of the lapse, $\beta$ is the
scalar part of the perturbation of the shift and $\gamma$ is a scalar
function. In this gauge the equations for the field perturbation can
be written as \citep{Lyth:2009zz}
\begin{eqnarray}
    \delta\ddot{\varphi}_{kI}+3H\delta\dot{\varphi}_{kI}+\frac{k^{2}}{a^{2}}\delta\varphi_{kI}+V_{,IJ}\delta\varphi_{k}^{J} & = & a^{-3}\frac{\mathrm{d}}{\mathrm{d}t}\left(\frac{a^{3}}{H}\dot{\varphi}_{I}\dot{\varphi}_{J}\right)\delta\varphi_{k}^{J}\,,\label{dFd-flat}
\end{eqnarray}
where repeated indices imply summation and for brevity we used the
notation $\varphi_{I}=\left(\phi,\chi\right)$ and similarly for the
perturbation. The potential $V\left(\phi,\chi\right)$ is provided
in Eq.~(\ref{V}). For the homogeneous value of the $\chi$ field,
we used the same expression as in Eq.~(\ref{chi-hom}).

Due to spatial homogeneity of the background FRW metric perturbation
variables $\beta$ and $\gamma$ come in the combination given by \citep{Hwang:1991aj}\footnote{The $\Psi$ symbol here should not be confused with the metric perturbation
in the Newtonian gauge in Eq.~(\ref{ds2-newt}).}
\begin{eqnarray}
\Psi_{k} & \equiv & a\left(\beta_{k}+a\dot{\gamma}_{k}\right)\,.\label{Psi-flat-def}
\end{eqnarray}
This variable satisfies the equation
\begin{eqnarray}
2H\frac{k^{2}}{a^{2}}\Psi_{k} & = & -\dot{\varphi}_{I}\delta\dot{\varphi}^{I}+\left(\ddot{\varphi}_{I}+\frac{\dot{\varphi}_{J}\dot{\varphi}^{J}}{2H}\dot{\varphi}_{I}\right)\delta\varphi_{k}^{I}\,.\label{Psi-flat}
\end{eqnarray}
In terms of these variables the curvature perturbation on the uniform
density slice $\zeta$ is given by 
\begin{eqnarray}
\zeta_{k} & = & -\frac{H\left(\dot{\varphi}_{I}\delta\varphi_{k}^{I}+\frac{2}{3}\frac{k^{2}}{a^{2}}\Psi_{k}\right)}{\dot{\varphi}_{I}\dot{\varphi}^{I}}\,.\label{z-flat}
\end{eqnarray}

In order to check our numerical computations we run the simulations
in the Newtonian and flat gauges independently and then check if they
give consistent results. To perform the comparison we need the expressions
that relate the various quantities in the two gauges. The final result for $\zeta$ can be checked by comparing Eq.~(\ref{z-flat}) above
with Eq.~(\ref{z-newt}). But we also compare intermediate quantities.
For example, the Newtonian curvature perturbation $\Phi_{k}$ in Eq.~(\ref{ds2-newt})
and $\Psi_{k}$ in Eq.~(\ref{Psi-flat-def}) above are related by
\begin{eqnarray}
\Phi_{k} & = & H\Psi_{k}\,.
\end{eqnarray}
We checked numerically that $\Phi_{k}$ computed solving Newtonian
gauge equations coincide exactly with the flat gauge solution of $\Psi_{k}$
after performing the conversion of the latter to $\Phi_{k}$ using
the above equation. Similarly, we can compare the scalar field perturbations. 

Similar conclusions hold for the field perturbations too. Such perturbations in the two
gauges are related by
\begin{eqnarray}
\left.\delta\varphi_{k}^{I}\right|_{\mathrm{Newt}} & = & \left.\delta\varphi_{k}^{I}\right|_{\mathrm{flat}}-\dot{\varphi}^{I}\Psi_{k}\,.
\end{eqnarray}
Again, the LHS of this equation, as computed from Eqs.~(\ref{dphi-eqn})
and (\ref{dchi-eqn}), coincides with the RHS, as
computed using Eqs.~(\ref{dFd-flat}) and (\ref{Psi-flat}).

\section{Detailed computations of induced gravitational waves}\label{app:sigw}
In this section, we introduce the detailed computations of the
spectrum of induced gravitational waves in the radiation-dominated
epoch~\cite{Espinosa:2018eve,Kohri:2018awv,Terada:2025cto}, which were
only summarized in Section~\ref{sec:sigw}.  Here we do not consider
non-gaussian perturbation for
simplicity~\cite{Cai:2018dig,Li:2023xtl,Li:2024zwx}. As given in
Eq.~(\ref{eq:sigw}), the spectrum of induced gravitational waves is
expressed to be
\begin{equation}\label{eq:sigw2}
  \Omega_{\mathrm{GW}}(\eta, k)=\frac{\rho_{\mathrm{GW}}(\eta,k)}{\rho_{\mathrm{tot}}(\eta)}=\frac{1}{24}\left(\frac{k}{a(\eta) H(\eta)}\right)^{2} \overline{\mathcal{P}_{T}(\eta, k)},
\end{equation}
where $\eta$ is the conformal time, the wave number $k$ is related to
the GW frequency $f$ via $k=2\pi f$, and $\rho_{\mathrm{GW}}(\eta,k)$ is
the energy density of gravitational waves per logarithmic wave
number. The overline denotes the oscillation
average. $\mathcal{P}_T(\eta, k)$ is the power spectrum of the tensor
perturbation $T$ which is expressed by
\begin{equation}\label{eq:ph2}
\mathcal{P}_{T}(\eta, k)=4 \int_{0}^{\infty} \mathrm{d} v \int_{|1-v|}^{1+v} \mathrm{~d} u\left(\frac{4 v^{2}-\left(1+v^{2}-u^{2}\right)^{2}}{4 v u}\right)^{2} I^{2}(v, u, x) \mathcal{P}_{\zeta}(k v) \mathcal{P}_{\zeta}(k u),
\end{equation}
where $x$ is the dimensionless variable $x = k\eta$. The variables $u$ and $v$ are defined by $u = |\mathbf{k} - \tilde{\mathbf{k}}| / k$ and $v = \tilde{k} / k$, respectively. The function $I(v, u, x)$ is an oscillating kernel function encoding the source information given by
\begin{equation}\label{eq:osc}
I(v, u, x)=\int_{0}^{x} \mathrm{~d} \bar{x} \frac{a(\bar{\eta})}{a(\eta)} k G_{k}(\eta, \bar{\eta}) f(v, u, \bar{x}),
\end{equation}
where $G_{k}$ is the Green's function, defined by the solution to the differential equation
\begin{equation}
  \label{eq:}
G_{k}''(\eta, \bar{\eta}) +\left( k^2 - { a''(\eta)}/{a(\eta)}\right) G_{k}(\eta, \bar{\eta}) = \delta (\eta - \bar{\eta}),
\end{equation}
with primes denoting derivatives with respect to $\eta$. The function
$f(v, u, \bar{x})$ represents the source term constructed from
second-order scalar perturbations. 
Since the kernel function $I$ with dissipation has been presented in Sec.~\ref{sec:sigw}, this Appendix provides the expression for the kernel function without dissipative effect, which is shown by the black dashed curve in Fig.~\ref{fig:sigw}.
To evaluate the spectrum of GWs observed at present, we take the late-time limit
$\eta \rightarrow \infty$, or equivalently $x \gg 1$. In addition, if
we take the oscillation average in this limit, we obtain
\begin{align}\label{eq:oscavg}
\nonumber
\overline{I_{\text{RD}}^2(v,u,x\to \infty)} &= \frac{1}{2} \left( \frac{3(u^2+v^2-3)}{4 u^3 v^3 x} \right)^2 \left( \left( -4uv+(u^2+v^2-3) \log \left| \frac{3-(u+v)^2}{3-(u-v)^2} \right| \right)^2 \right. \\
&\quad\left. + \pi^2 (u^2+v^2-3)^2 \Theta ( v+u-\sqrt{3}) \right),
\end{align}
where $\Theta$ denotes the Heaviside theta function. Furthermore, as pointed 
out in Ref.~\cite{Kohri:2018awv}, due to the symmetry under the exchange of $u$ 
and $v$, we can perform a change of variables from $(u,v)$ to $(t,s)$ with 
$t = u + v - 1$ and $s = u - v$. 
Under this transformation, the oscillation-averaged kernel function becomes
\begin{align}
\label{eq:oscts}
\nonumber 
\overline{I_{\text{RD}}^2(t,s,x\rightarrow \infty)} =& \frac{288(-5+s^2+t(2+t))^2}{x^2 (1-s+t)^6 (1+s+t)^6}\left( \frac{\pi^2}{4} (-5+s^2+t(2+t))^2 \Theta (t -(\sqrt{3}-1)) \right. \\
& \left.+ \left( -(t-s+1)(t+s+1) + \frac{1}{2} (-5+s^2+t(2+t)) \log \left| \frac{-2+t(2+t)}{3 - s^2} \right| \right)^2 \right).
\end{align}
Rewriting the power spectrum of the tensor perturbations
$\mathcal{P}_T(\eta, k)$ in terms of the new variables $s$ and $t$,
one obtains
\begin{equation}\label{eq:phst}
\mathcal{P}_{T}(\eta, k)=2 \int_{0}^{\infty} \mathrm{d} t \int_{-1}^{1} \mathrm{~d} s\left[\frac{t(2+t)\left(s^{2}-1\right)}{(1-s+t)(1+s+t)}\right]^{2} I^{2}(t, s, x\rightarrow \infty) \mathcal{P}_{\zeta}\left(k \frac{t-s+1}{2}\right) \mathcal{P}_{\zeta}\left(k \frac{t+s+1}{2}\right).
\end{equation}

\section{Detailed computations of gravitational waves from merging binary PBHs}\label{app:gwmerge}

\begin{figure}
\begin{centering}
\includegraphics[width=10cm]{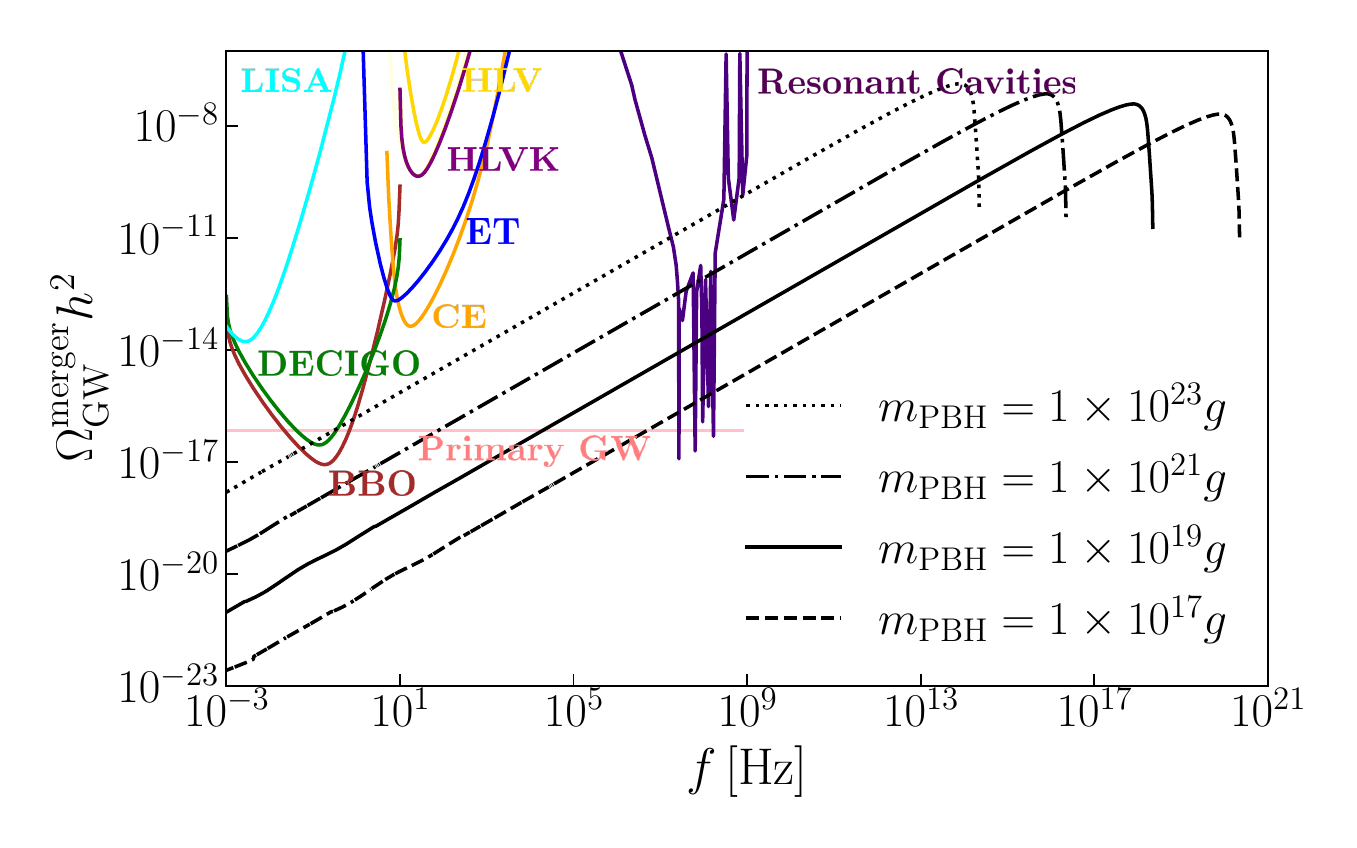}
\par\end{centering}
\caption{
    GW spectrum generated by PBH mergers. 
    The horizontal axis represents the frequency $f$ in units
    of Hz, while the vertical axis corresponds to the energy density of
    the gravitational waves originating from PBH mergers, denoted by
    $\Omega_{\text{GW}}^{\text{merger}} h^{2}$, where $h$ is the
    dimensionless Hubble parameter. 
    The black curve represents the spectrum of the gravitational waves from mergers of PBH with
    $10^{17}\text{g}$(dashed), $10^{19}\text{g}$(solid), 
    $10^{21}\text{g}$(dot-dashed), and $10^{23}\text{g}$(dashed) respectively within the sensitivity range of resonant cavities of axion. Furthermore, the signal from PBHs with $10^{23}\text{g}$
    (dotted line) can be observed within the sensitivity of reach not only
    the resonant cavities of axion but also DECIGO and BBO.
    The pink curve approximately corresponds to the spectrum of primary gravitational waves. 
    The other colored curves indicate the sensitivity curves of future planned gravitational-wave
    observatories~\cite{Schmitz:2020syl}, and resonant cavities for the axion detection~\cite{Berlin:2021txa,Herman:2022fau}.
}
\label{fig:gwmergg}
\end{figure}

In this section, we discuss the details of the computations of the
gravitational wave spectrum emitted from merging binary PBHs 
~\cite{Wang:2019kaf,Kohri:2024qpd,Sasaki:2018dmp} which was
only summarized in Section~\ref{sec:gwmerg}. The spectrum of
gravitational waves from the PBH mergers is given by
Eq.~(\ref{eq:gwmerg}), which we write here again
\begin{equation}\label{eq:gwmerg2}
\Omega_\text{GW}^\text{(merger)}(f) = \frac{f}{\rho_\text{c}} \int_0^{z_\text{sup}} \mathrm{d} z \, \frac{R(z)}{(1 + z) H(z)} \frac{\mathrm{d}E(f_\text{s})}{\mathrm{d}f_\text{s}},
\end{equation}
where the Hubble parameter is given by
$H(z)=H_{0}[\Omega_{\text{r}}(1+z)^{4}+\Omega_{\text{m}}(1+z)^{3}+\Omega_{\Lambda}]^{1/2}$with
$\Omega_{\text{r}}$ and
$\Omega_{\Lambda} = 1 - \Omega_{\text{r}} - \Omega_{\text{m}}$ being
the $\Omega$ parameters of radiation, and the present-day
cosmological constant, respectively. The quantity $R(z)$ denotes the
rate of the mergers for the binary PBHs per comoving volume for a PBH
mass $m_{\mathrm{PBH}}$, given by
\begin{equation}\label{eq:mrate}
R(z) = \frac{f_\text{PBH} \Omega_\text{CDM} \rho_\text{c}}{m_{\mathrm{PBH}}} \frac{\mathrm{d} P_t}{\mathrm{d} t},
\end{equation}
where ${\mathrm{d} P_t}/{\mathrm{d} t}$ is the probability distribution for a PBH merger occurring at time $t$, given by
\begin{equation}\label{eq:dPdt}
   \frac{\mathrm{d} P_t}{\mathrm{d} t} = \frac{3}{58 t} \times \begin{cases}
       \left(\frac{t}{T_{\rm per}} \right)^{3/37} - \left( \frac{t}{T_{\rm per}} \right)^{3/8} & (t < t_\text{c}) \\
       \left(\frac{t}{T_{\rm per}}\right)^{3/8} \left( \left(\frac{t}{t_\text{c}}\right)^{-29/56}\left( \frac{4\pi f_\text{PBH}}{3} \right)^{-29/8} - 1\right) & (t \geq t_\text{c})
   \end{cases}, 
\end{equation}
with $t_\text{c}$ defined by $t_\text{c} = (4\pi f_\text{PBH}/3)^{37/3} T_{\rm per}$ and $T_{\rm per}$ given by
\begin{equation}\label{eq:lt}
T_{\rm per} = \frac{729}{340 \pi^2 (1 + z_\text{eq})^4 (4 \pi f_\text{PBH}^{16} m_{\mathrm{PBH}}^5 \rho_\text{c}^4 / 3)^{1/3}},
\end{equation}
where $z_\text{eq}$ is the redshift at the epoch of matter-radiation equality. The typical time $t$ at which mergers occur is given by
\begin{equation}\label{eq:time}
t=\int_{z}^{\infty} \frac{dz^{\prime}}{(1+z^\prime)H(z^\prime)},
\end{equation}
In Eq.~(\ref{eq:gwmerg}), the GW energy spectrum of a non-spinning PBH binary from the source, $\mathrm{d}E(f_\text{s})/\mathrm{d}f_\text{s}$, is modeled as
\begin{equation}\label{eq:dedf}
\frac{\mathrm{d}E(f_\text{s})}{\mathrm{d}f_\text{s}}=\frac{(G\pi)^{2/3}M_{c}^{5/3}}{3}
\begin{cases}
f_{\mathrm{s}}^{ -1/3} & \text{for}~f_{\mathrm{s}}<f_1; \; \text{inspiral phase} \\
w_1 f_{\mathrm{s}}^{2/3} & \text{for}~f_1\le f_{\mathrm{s}}<f_2; \; \text{merger phase} \\
w_2\frac{f_{\mathrm{s}}^{2}}{\left(1+\frac{4(f_{\mathrm{s}}-f_{2})^{2}}{\sigma^{2}}\right)^{2}} & \text{for}~f_2\le f_{\mathrm{s}}\le f_3; \; \text{ringdown phase} \\
0, & \text{for}~f_3 < f_{\mathrm{s}}
\end{cases}
\end{equation}
where $f_\mathrm{s} = (1+z)f$ denotes the frequency emitted at the source, $G$ is the Newton’s gravitational constant and $M_{c}$ is the chirp mass defined by $M_{c}^{5/3} = m_\text{PBH,1}m_\text{PBH,2}/(m_\text{PBH,1}+m_\text{PBH,2})^{1/3}$. The parameters $w_1$ and $w_2$ are fitting coefficients chosen to ensure the continuity of the spectrum, given by $w_1 = f_1^{-1}$ and $w_2 = f_1^{-1} f_2^{-4/3}$. $f_1$, $f_2$, $f_3$, and $\sigma$ are given
\begin{eqnarray}\label{eq:f123}
&&\pi M_{t} f_1 = (1-4.455+3.521)+0.6437\eta-0.05822\eta^2-7.092\eta^3\\
&&\pi M_{t} f_2 = (1-0.63)/2+0.1469\eta-0.0249\eta^2+2.325\eta^3\\
&&\pi M_{t} f_3 = 0.3236 -0.1331\eta -0.2714\eta^2 +4.922\eta^3\\
&&\pi M_{t} \sigma = (1-0.63)/4 -0.4098\eta +1.829\eta^2-2.87\eta^3,
\end{eqnarray}
where the total mass $M_t = m_\text{PBH,1} + m_\text{PBH,2}$ and the symmetric mass ratio $\eta = m_\text{PBH,1} m_\text{PBH,2} / (m_\text{PBH,1} + m_\text{PBH,2})^2$.

We now consider the asteroid-mass range
($10^{17} \mbox{g} \lesssim m_{\text{PBH}} \lesssim 10^{23}
\mbox{g}$),
in which an abundance $f_{\text{PBH}} \sim \mathcal{O}(1)$ is allowed
under observational constraints to become the 100$\%$ CDM. Within this
mass range, we choose four benchmark values, $10^{17}\text{g}$,
$10^{19}\text{g}$, $10^{21}\text{g}$, and $10^{23}\text{g}$, and
compute the corresponding spectra of the gravitational waves from the PBH
mergers. These plots are shown as the black curves in
Fig.~\ref{fig:gwmergg}. 

\bibliographystyle{apsrev4-2}
\bibliography{rmitt-citations}

\end{document}